\documentstyle[12pt, epsf]{article}
\input psfig.tex
\begin{document}

\def\ov{\overline}
\def\ra{\rightarrow}
\def\epslash{\not{\hbox{\kern -1.5pt $\epsilon$}}}
\def\pslash{\not{\hbox{\kern -1.5pt $p$}}}
\def\aslash{\not{\hbox{\kern -1.5pt $a$}}}
\def\bslash{\not{\hbox{\kern -1.5pt $b$}}}
\def\Dslash{\not{\hbox{\kern -4pt $D$}}}
\def\wslash{\not{\hbox{\kern -4pt $\cal W$}}}
\def\zslash{\not{\hbox{\kern -4pt $\cal Z$}}}
\def\kln{\kappa_{L}^{NC}}
\def\krn{\kappa_{R}^{NC}}
\def\klc{\kappa_{L}^{CC}}
\def\krc{\kappa_{R}^{CC}}
\def\bbz{{\mbox {\,$b$-${b}$-$Z$}\,}}
\def\ttz{{\mbox {\,$t$-${t}$-$Z$}\,}}
\def\tta{{\mbox {\,$t$-${t}$-$A$}\,}}
\def\bba{{\mbox {\,$b$-${b}$-$A$}\,}}
\def\tbw{{\mbox {\,$t$-${b}$-$W$}\,}}
\def\Tr{{\rm Tr}}

\setcounter{footnote}{1}
\renewcommand{\thefootnote}{\fnsymbol{footnote}}


\begin{titlepage}

hep-ph/9606397 \hfill {\small MSUHEP-60620}
\begin{flushright}
{ June 20, 1996}
\end{flushright}

\vspace*{1.2cm}
\vspace*{0.5cm} 
\centerline{\Large\bf  
Top Quark Interactions and the}
\baselineskip=18pt
\centerline{\Large\bf  
Search for New Physics  }

\vspace*{1.2cm}
\baselineskip=17pt
\centerline{\normalsize  
{\bf F. Larios\footnote{Also at the 
Departamento de F\'{\i}sica, CINVESTAV, Apdo. Postal 14-740, 
07000 M\'exico, D.F., M\'exico.}~~ and~~ {\bf C.--P. Yuan} } }
 
\vspace*{0.4cm}
\centerline{\normalsize\it
Department of Physics and Astronomy, Michigan State University }
\centerline{\normalsize\it
East Lansing, Michigan 48824 , USA }

\vspace{0.4cm}
\raggedbottom
\setcounter{page}{1}
\relax

\begin{abstract}
\noindent
A complete characterization of all the possible effective
interactions up to dimension 5 of the top and bottom quarks
with the electroweak gauge bosons have been made within
the context of the non-linear chiral Lagrangian.
The dimension 5 operators (19 in total) can contribute to the
$V_L V_L \ra t\ov t$ or $t\ov b$ amplitudes with a leading
energy power of $E^3$ ($E^2$) for fermion pairs
with same (opposite) sign helicities. 
Because of the equivalence between the longitudinal weak
boson and the corresponding would-be Goldstone boson in
high energy collisions (Goldstone Equivalence Theorem),
they are sensitive to the electroweak symmetry breaking sector.
We also show the top quark production rates at the
LHC and the LC via  $V_L V_L$ fusion processes. 
At the LC, if no anomalous production rate is found, these
coefficients can be bound (based on the naive dimensional
analysis) to be of order $10^{-1}$.  This is about an order of
magnitude more stringent than the bounds for the
next-to-leading order bosonic operators commonly studied
in $V_L V_L \ra V_L V_L$ scatterings. 
The effects on a CP-odd observable are also briefly discussed.
\end{abstract}

\vspace*{3.4cm}
\end{titlepage}
\normalsize\baselineskip=15pt
\setcounter{footnote}{0}
\renewcommand{\thefootnote}{\arabic{footnote}}

\newpage

\section{Introduction}
\indent

Despite the unquestionable significance of its achievements,
like that of predicting the existence of the top quark, 
there is no reason to believe that the Standard Model (SM) is
the final theory.  For instance, the SM contains many arbitrary
parameters with no apparent connections.  In addition,  the SM
provides no satisfactory explanation for the symmetry-breaking
mechanism which takes place and gives rise to the observed mass
spectrum of the gauge bosons and fermions.   Because the top quark
is heavy relative to other observed fundamental particles, one expects
that any underlying theory, to supersede the SM at some high energy
scale $\Lambda \gg m_t$, will easily reveal itself at lower energies
through the effective interactions of the top quark to other light
particles.  Also because the top quark mass
($\sim {v/ {\sqrt{2}}}$) is of the order of the Fermi scale
$v={(\sqrt{2}G_F)}^{-1/2}=246$\,GeV \cite{topdisc},
which characterizes the electroweak symmetry-breaking scale,
the top quark system could be a useful probe for the
symmetry-breaking sector.  Furthermore, the fermion mass generation
can be closely related to the electroweak symmetry-breaking
mechanism, one expects some residual effects of this mechanism to
appear in accordance with the mass hierarchy
\cite{pczh,sekh,malkawi}. This means that new effects should
be more apparent in the top quark sector than in any other light
sector of the theory.  Therefore, it is important to study the top quark
system as a direct tool to probe new physics effects \cite{kane}.

Because of the great diversity of models proposed for possible 
new physics (beyond the SM), it has become necessary
to study these possible new interactions in a model independent
approach \cite{effec}. This approach has proved to render relevant
non-trivial information about the possible deviations from the standard
couplings of the heavier elementary particles (heavy scalar  bosons,
the bottom and the top quarks, etc.) \cite{miguel}.
Our study focuses on  the top quark, which
because of its remarkably higher mass is the best candidate
among the fermion particles for manifesting these anomalous
interactions at high energies.

A common approach to study these anomalous couplings is to 
consider the most general on-shell vertices (form factors) 
 involving the bottom and the top quarks 
and the gauge bosons of interest \cite{kane}. 
In this work we will incorporate the effective 
chiral Lagrangian approach \cite{chiral},  
which is based on the principle of gauge symmetry, 
but the symmetry is realized in the most general 
(non-linear) form so as to encompass all the possible interactions 
consistent with the existing experimental data.  
The idea of using this approach 
is to exploit the linearly realized ${\rm {U(1)}_{em}}$ symmetry
and the non-linearly realized $\rm{SU(2)}_L \times \rm{U(1)}_Y$
symmetry to make a systematic characterization of all the
anomalous couplings.  In this way, for example, different couplings
which otherwise would be considered as independent become
related through the equations of motion.  

In Ref.~\cite{malkawi} it was shown that low energy data
(including $Z$ pole physics) generally do not impose any stringent
constraints on the $\kappa$ coefficients of the anomalous
couplings in ${\cal L}^{(4)}$ [cf.  Eq.~(\ref{eq2})].  
Hence these anomalous couplings
 have to be directly measured via production of top 
quarks at the colliders.
For instance, the couplings ${\kappa}^{CC}_{L,R}$ can be
measured from the decay of the top quarks in $t\overline {t}$ pairs
produced either at hadron colliders, such as the Tevatron and
the Large Hadron Collider (LHC), or at the 
electron linear collider (LC).    They can also be studied from the 
production of the single-top quark events via 
$W$-gluon fusion ($W g \ra t \bar b$)
\cite{dawson,sally,wgfusion,toptalk,nlowg} or 
the Drell-Yan like ($W^{*} \;\ra t\ov {b}$)  \cite{wstar,nlowstar} 
processes at the hadron colliders, as well as
from the $W$-photon fusion ($W \gamma \ra t \bar b$) 
process at the electron colliders \cite{booslc}.
The coupling ${\kappa}^{NC}_{L,R}$ can only 
be sensitively probed at a future linear collider via the 
$e^{+}\;e^{-}\;\ra \gamma , Z \;\ra t\ov {t}$ process \cite{ladi}
because at hadron colliders the $t\ov {t}$ production rate is
dominated by QCD interactions ( $q\ov {q}, gg\;\ra\; t\ov {t}$ ). 
At the LHC ${\kappa}^{NC}_{L,R}$ may also be studied via the
associated production of $t\ov {t}$ with $Z$ bosons,
which deserves a separate study.

In this work, we will extend the previous study by including
dimension 5 fermionic operators, and then examine the precision
with which the coefficients of these operators can be measured in
high energy collisions.   Since it is the electroweak symmetry
breaking sector that we are interested in, we shall concentrate on
the interaction of the top quark with the longitudinal weak gauge
bosons; which are equivalent to the would-be Goldstone
bosons in the high energy limit.  This equivalence is known
as the Goldstone Equivalence Theorem  \cite{et1pol}-\cite{et}. 

For simplicity, we will only construct the complete set of dimension 5 
effective operators for the fermions $t$ and $b$, although our results
can be trivially extended for operators 
including other fermions such as the flavor changing 
neutral interactions $t$-$c$-$Z$, $t$-$c$-$\gamma$ \cite{tcz}
and $t$-$c$-$g$ \cite{tcg}.

Our strategy for probing these anomalous dimension 5 operators 
( ${\cal L}^{(5)}$ ) is to study the production of $t \ov t$ pairs as
well as single-$t$ or $\ov t$ via the $W^+_L W^-_L$, $Z_L Z_L$
and $W^\pm_L Z_L$ (denoted in general as $V_L V_L$) fusion
processes in the TeV region. 
As we shall show later, the leading contribution of the 
scattering amplitudes at high energy goes as $E^3$ for the
anomalous operators  ${\cal L}^{(5)}$, 
where $E\,=\,\sqrt{s}$ is the CM energy of the $W^+ W^-$
or $Z Z$ system (that produces $t\ov t$), or the $W^\pm Z$
system (that produces $t\ov b$ or $b\ov t$). 
On the other hand,  when the
coefficients $\kappa^{CC}_{L,R}$ and $\kappa^{NC}_{L,R}$ are zero,
the dimension 4 operators ${\cal L}^{(4)}$ can at most contribute with
the first power $E^1$ to these scattering $V_L V_L$ processes.
Hence, in this case, the 
$V_L V_L \ra f \ov f'$ scatterings in the high energy region are more
sensitive to ${\cal L}^{(5)}$ than to ${\cal L}^{(4)}$.  
If these $\kappa$'s are not zero, then the high energy behavior can
at most grow as $E^2$ as compared to $E^3$
for the dimension 5 operators (See Appendix B).

As mentioned before, the dimension 4 anomalous couplings
$\kappa$'s are better measured at the
scale of $M_W$ or $m_t$ by  
studying the decay or the production of the top quark at either
the Tevatron, the LHC, or the LC near the $t\ov t$ threshold.
Since, as mentioned above, the dimension 5 operators are
better measured in the TeV region, we shall assume that by the
time their measurement is feasible, the $\kappa$'s will already be
known.  Thus, to simplify our discussion,
we will take the values of the $\kappa$'s to be
zero when presenting our numerical results.

We show that there are 19 independent dimension 5 
operators (with only $t$, $b$ and gauge boson fields)
in ${\cal L}^{(5)}$ after imposing the equations of motion for the
effective chiral Lagrangian.  It is expected that at
the LHC or the LC there will be  about a few hundreds to
a few thousands of $t\ov t$ pairs or single-$t$
(or single-$\ov t$) events produced via the $V_L V_L$ 
fusion processes.
The coefficients of these operators, with the pre-factor 
$\frac {1}{\Lambda}$, could be measured at the LC (less
likely at the LHC) to order  $10^{-2}$ or $10^{-1}$.
As will shown later, the pre-factor $\frac {1}{\Lambda}$ is 
suggested by the naive dimensional analysis \cite{georgi}, 
and $\Lambda$ is 
the cut-off scale of the effective theory.  It could be the lowest new 
heavy mass scale, or something around $4 \pi v \simeq 3.1$
TeV if no new resonances exist below $\Lambda$.
As a comparison, the coefficients of the NLO 
bosonic operators  are usually determined to about an order
of $10^{-1}$ or $1$ via $V_LV_L \ra V_LV_L$
processes \cite{et,sss}.  Hence, the scattering processes
 $V_L V_L \ra t\ov t, t\ov b, \,{\rm or}\, b \ov t$ 
at high energy may be more sensitive for probing some 
symmetry breaking mechanisms than $V_LV_L \ra V_LV_L$.

This paper is organized as follows.  In section 2 the general
framework of the Electroweak Chiral Lagrangian is presented. 
In section 3 we make a systematic characterization of all the
independent dimension 5 operators that are invariant under the
symmetry of the gauge group.    Section 4 deals 
with the CP properties of these interactions.   In section 5 we make
a general analysis of their potential contribution to the production
cross section of top quarks according to their behavior at high
energies.  A very useful simplification is made by considering an
approximate custodial symmetry in our set of operators;
this is discussed in section 6.       
Section 7 contains the analytical 
results for the amplitudes of various $V_L V_L$ 
fusion processes; an approximate custodial symmetry is assumed
here, but the general expressions can be found in
Appendices B and C.    Finally, in section 8 we discuss
the values for the coefficients of these anomalous 
operators at which a significant signal can appear at 
both the LHC and the LC; also, we include an example of how
one can measure possible CP violating effects coming from
these operators.  Section 9 has our conclusions.

\section{The ingredients of the Electroweak Chiral Lagrangian}
\indent

We consider the electroweak theories in which the gauge symmetry 
$G \equiv {\rm{SU(2)}}_{L}\times {\rm{U(1)}}_{Y}$ is spontaneously
broken down to $H={\rm {U(1)}_{em}}$\cite{malkawi,georgi,chan,app}. 
There are three Goldstone bosons, $\phi^{a}$ ($a=1,2,3$), generated
by this breakdown of $G$ into $H$, which are eventually {\it eaten} by
the $W^{\pm}$ and $Z$ gauge bosons and become their longitudinal
degrees of freedom.

In the non-linearly realized chiral Lagrangian formulation, the
Goldstone bosons transform non-linearly under $G$ but linearly
under the subgroup $H$. A convenient way to handle this is to
introduce the matrix field
\begin{equation}
\Sigma ={\rm{exp}}\left( i\frac{\phi^{a}\tau^{a}}{v_{a}} \right)
\label{sigfield}\, ,
\end{equation}
where $\tau^{a},\, a=1,2,3$ are the Pauli matrices normalized as
${\rm{Tr}}(\tau^a \tau^b)=2 \delta_{ab}$.
The matrix field $\Sigma$ transforms under $G$ as
\begin{equation}
\Sigma\ra {\Sigma}^{\prime}=\,g_L \Sigma \,g^\dagger_R\, ,
\end{equation}
with
\begin{eqnarray}
g_L =&& {\rm {exp}}\left ( i\frac{\alpha^{a}\tau^{a}}{2}\right )\; ,\\
g^\dagger_R =&& {\rm {exp}}(-i\frac{y\tau^3}{2})\; ,\nonumber
\end{eqnarray}
where $\alpha^{1,2,3}$ and $y$ are
the group parameters of $G$.
Because of the ${\rm {U(1)}_{em}}$ invariance,  
$v_1=v_2=v$ in Eq.~(\ref{sigfield}), 
but they are not necessarily equal to $v_3$.
In the SM, $v$ is the vacuum expectation value of the Higgs
boson field, and characterizes the scale of the symmetry-breaking.
Also, $v_3=v$ arises from the approximate custodial symmetry
present in the SM.
It is this symmetry that is responsible for the
tree-level relation
\begin{equation}
\rho= \frac{M_W^2}{M_Z^2\,\cos^2 \theta_W}=1\,
\label{yeq3}
\end{equation}
in the SM, where $\theta_W$ is the electroweak mixing angle,
$M_W$ and $M_Z$ are the masses of $W^\pm$ and $Z$ boson,
respectively.  In this paper we assume the underlying theory
guarantees that $v_1=v_2=v_3=v$.

In the context of this non-linear formulation of the electroweak
theory, the massive charged and neutral weak bosons can be
defined by means of the {\it composite} field:
\begin{equation}
{{\cal W}_{\mu}^a}=-i{\rm Tr}(\tau^{a}\Sigma^{\dagger}
D_{\mu}\Sigma)\,
\end{equation}
where\footnote{This is not the covariant derivative of $\Sigma$. 
The covariant derivative is \\
$D_{\mu}\Sigma=\partial_{\mu}\Sigma-ig\frac{\tau^a}
{2}W_{\mu}^a \Sigma+ig^{\hspace{.5mm}\prime}\Sigma
\frac{\tau^3}{2}B_{\mu}$, such that $D_\mu \Sigma\ra D_\mu
{\Sigma}^{\prime}=\,g_L (D_\mu \Sigma )\,g^\dagger_R$.}
\begin{equation}
D_{\mu}\Sigma=\left (\partial_{\mu}-ig\frac{\tau^a}{2}
W_{\mu}^a\right) \Sigma\,\,\, .\label{covderw}
\end{equation}
Here, $W_\mu^a$ is the gauge boson associated with the 
${\rm{SU(2)}}_L$ group, and its transformation is 
\begin{equation}
\tau^a W_\mu^a \ra \tau^a W_\mu^{'a} \;=\;
 g_L\; \tau^a W_\mu^a\; g_L^{\dagger}\;+\;
\frac{2i}{g} g_L\partial_\mu g_L^{\dagger}\, ,
\end{equation}
where $g$ is the gauge coupling. 
The $D_{\mu}\Sigma$ term transforms under $G$ as
\begin{equation}
D_{\mu}\Sigma \ra D_{\mu}\Sigma^{'} = 
g_L \left( D_{\mu}\Sigma \right) g^\dagger_R +
g_L \Sigma \partial_\mu g^\dagger_R \; .
\end{equation}
Therefore, by using the commutation rules for the Pauli matrices
and the fact that $\Tr(AB)=\Tr(BA)$ we can prove that the
composite field ${{\cal W}_{\mu}^a}$ will transform under $G$
in the following manner:
\begin{equation}
 {{\cal W}_{\mu}^3}\rightarrow {{{\cal W}^{\prime}}_{\mu}^3}
       ={{\cal W}_{\mu}^3}-\partial_{\mu}y\, ,
\end{equation}
\begin{equation}
{{\cal W}_{\mu}^\pm}\rightarrow {{{\cal W}^{\prime}}_{\mu}^\pm}
  =e^{\pm iy}{{\cal W}_{\mu}^\pm} \label{wtransf}\; ,
\end{equation}
where
\begin{equation}
{\cal W}_{\mu}^{\pm}={\frac{{\cal W}_{\mu}^{1}\mp
i{\cal W}_{\mu}^{2}}
{\sqrt{2}}}\, .\label{wdef}
\end{equation}
Also, it is convenient to define the field 
\begin{equation}
{\cal B}_{\mu}=g^{\hspace{.5mm}\prime}B_{\mu}\,\, ,
\label{b}
\end{equation}
which is really the same gauge boson field associated with the
${\rm{U(1)}}_Y$ group ($g^{\hspace{.5mm}\prime}$ is the gauge
coupling). The field ${\cal B}_{\mu}$ transforms under $G$ as
\begin{equation}
{\cal B}_{\mu} \rightarrow {\cal B}^{\prime}_{\mu} =
 {\cal B}_{\mu}+\partial_{\mu}y\,
\label{bb}\, .
\end{equation}

We now introduce the composite fields
${\cal Z}_{\mu}$ and ${\cal A}_{\mu}$ as
\begin{equation}
{\cal Z}_\mu={\cal W}^3_\mu +{\cal B}_\mu
\label{b1}\,\, ,
\end{equation}
\begin{equation}
s_w^2{\cal A}_\mu = s_w^2{\cal W}^3_\mu - c_w^2 {\cal B}_\mu\, ,
\label{b2}
\end{equation}
where $s_w^2\equiv\sin^2\theta_W$, and $c_w^2=1-s_w^2$.
In the unitary gauge ($\Sigma =1$)
\begin{equation}
{\cal W}_{\mu}^a=-gW_{\mu}^a \,\, ,
\end{equation}
\begin{equation}
{{\cal Z}}_{\mu} =-\frac{g}{c_w} Z_{\mu}\,\, ,
\end{equation}
\begin{equation}
{{\cal A}}_{\mu}=-\frac{e}{s_w^2}A_{\mu} \label{afield}\,\, ,
\end{equation}
where we have used the relations
$e=g s_w=g^{\hspace{.5mm}\prime} c_w$,
$W_\mu^3= c_w Z_\mu + s_w A_\mu$, and
$B_\mu= -s_w Z_\mu + c_w A_\mu$.
In general, the composite fields contain Goldstone boson fields:
\begin{eqnarray}
{{\cal Z}}_{\mu}\; =&&
-\, \frac{g}{c_w} Z_{\mu} \; + \; {2\over {v}} {\partial}_{\mu}
{\phi}^3 \; -\; i \frac{2 g}{v} (W^{+}_{\mu} {\phi}^{-} -W^{-}_{\mu}
{\phi}^{+}) \; + \; \nonumber \\
&& i \frac{2}{v^2} ({\phi}^{-} {\partial}_{\mu} {\phi}^{+} - 
{\phi}^{+} {\partial}_{\mu} {\phi}^{-}) \;
+ \; \cdot \cdot \cdot\; , \label{zexp} \\  
{{\cal W}}_{\mu}^{\pm}\; =&&- g W_{\mu}^{\pm} \; + \;
{2\over {v}}\partial_{\mu} {\phi}^{\pm} \; \pm \; 
i \frac{2 g}{v} ({\phi}^{3} W^{\pm}_{\mu}  - W^{3}_{\mu}
{\phi}^{\pm}) \; \pm \; \nonumber \\
&& i \frac{2}{v^2} ({\phi}^{\pm} {\partial}_{\mu} {\phi}^{3} - 
{\phi}^{3} {\partial}_{\mu} {\phi}^{\pm}) \;
+\cdot \cdot \cdot\; , \label{wexp} 
\end{eqnarray}
where $\cdot \cdot \cdot$ denotes terms with 3 or more
boson fields.

The transformations of ${\cal Z}_{\mu}$ and
${\cal A}_{\mu}$ under $G$ are
\begin{equation}
{\cal Z}_{\mu}\rightarrow {\cal Z}_{\mu}^{\prime}=
{\cal Z}_{\mu}\label{ztransf}\, ,
\end{equation}
\begin{equation}
{\cal A}_{\mu} \rightarrow {\cal A}_{\mu}^{\prime} =
{\cal A}_{\mu} -\frac{1}{s_w^2}\partial_{\mu}y\label{atransf} \,\, .
\end{equation}
Hence, under $G$ the fields ${\cal W}_\mu^\pm$ and ${\cal Z}_\mu$
transform as vector fields, but ${\cal A}_\mu$ transforms as a gauge
boson field which plays the role of the photon field $A_\mu$.

Using the fields defined as above, one may construct
the ${\rm{SU(2)}}_L \times {\rm{U(1)}}_Y$ gauge invariant interaction
terms in the chiral Lagrangian
\begin{eqnarray}
{\cal L}^B =&-&\frac{1}{4g^2} {{\cal W}_{\mu \nu}^a}
{{\cal W}^{a}}^{\mu \nu}
 -\frac{1}{4{g^\prime}^2} {\cal B}_{\mu \nu}{\cal B}^{\mu \nu}
\nonumber \\
&+&\frac{v^2}{4}{\cal W}^{+}_{\mu}{{\cal W}^{-}}^{\mu}+
\frac{v^2}{8} {\cal Z}_{\mu}{\cal Z}^{\mu}+{\dots }\,\, ,
\label{eq4}
\end{eqnarray}
where
\begin{equation}
{\cal W}^{a}_{\mu \nu}=\partial_{\mu}{\cal W}^{a}_{\nu}
-\partial_{\nu}{\cal W}^{a}_{\mu}+\epsilon^{abc}
{\cal W}^{b}_{\mu} {\cal W}^{c}_{\nu}\label{wstrength} \,\, ,
\end{equation}
\begin{equation}
{\cal B}_{\mu \nu}=\partial_{\mu}{\cal B}_{\nu}-\partial_{\nu}
{\cal B}_{\mu}\,\, ,
\end{equation}
and where ${\dots}$ denotes other possible four-
or higher-dimension operators \cite{app}.
It is easy to show that\footnote{
Use ${\cal W}_{\mu}^a \tau^a = -2 i \Sigma^{\dagger} D_\mu
\Sigma ~$, and $[\tau^a,\tau^b]=2 i \epsilon^{abc} \tau^c $.}
\begin{equation}
{\cal W}_{\mu \nu}^a \tau^a=-g\Sigma^{\dagger}
W^a_{\mu \nu}\tau^a \Sigma\,\,
\end{equation}
and
\begin{equation}
{{\cal W}_{\mu \nu}^a} {{\cal W}^{a}}^{\mu \nu}=g^2
W_{\mu \nu}^{a} {{W^a}^{\mu \nu}}\,\, ,
\label{eq05}
\end{equation}
which does not have any explicit dependence on $\Sigma$.
This simply reflects the fact that the kinetic term is not related to
the Goldstone bosons sector, i.e. it does not originate from the
symmetry-breaking sector.

The mass terms in Eq.~(\ref{eq4}) can be expanded as
\begin{eqnarray}
\frac{v^2}{4}{\cal W}_{\mu}^{+}{{\cal W}^{-}}^{\mu}
+\frac{v^2}{8}{\cal Z}_{\mu}{{\cal Z}}^{\mu}
&=&{\partial}_{\mu}\phi^{+}\partial^{\mu}\phi^{-}
+\frac{1}{2}\partial_{\mu}\phi^{3}\partial^{\mu}\phi^{3}
\nonumber \\
&&+\frac{g^2v^2}{4}W_{\mu}^{+}{W^{\mu}}^{-}
+\frac{g^2v^2}{8c_w^2}Z_{\mu}Z^{\mu}+{\dots}\,\, .
\end{eqnarray}
At the tree-level, the mass of $W^\pm$ boson is $M_W=gv/2$ and
the mass of $Z$ boson is $M_Z=gv/2c_w$.

Fermions can be included in this context by assuming that each flavor
transforms under $G={\rm{SU(2)}}_L\times {\rm{U(1)}}_{Y}$ as
\begin{equation}
f\rightarrow {f}^{\prime}=e^{iyQ_f}f \label{eq1} \, ,
\end{equation}
where $Q_{f}$ is the electric charge of $f$.
($Q_f=\frac{2}{3}$ for the top, and $Q_f=-\frac{1}{3}$ for the
bottom quark.)

Out of the fermion fields $f_1$, $f_2$ (two different flavors)
and the Goldstone bosons matrix field $\Sigma$,
the usual linearly realized fields
$\Psi$ can be constructed. For example, the left-handed
fermions [${\rm{SU(2)}}_L$ doublet] are
\begin{equation}
\Psi_{L} \equiv {\psi_1\choose {\psi_2}}_{L} \; = \Sigma F_{L}\; = 
\Sigma{f_1\choose {f_2}}_{L} \label{psi} \,
\end{equation}
with $Q_{f_1}-Q_{{f_2}}=1$.
One can easily show that $\Psi_{L}$\,transforms linearly
under $G$ as
\begin{equation}
\Psi_{L}\rightarrow {\Psi}^{\prime}_{L}={\rm g} \Psi_{L}\, ,
\end{equation}
where ${\rm{g}}=
{{\rm {exp}}}(i\frac{\alpha^{a}\tau^{a}}{2})
{{\rm {exp}}}(i y\frac {Y}{2})\in G $, and 
$Y\;=\;\frac{1}{3}$ is the hypercharge of the left handed quark doublet.

In contrast, linearly realized right-handed fermions
$\Psi_{R}$  [${\rm{SU(2)}}_L$ singlet] simply coincide
with $F_{R}$, i.e.,
\begin{equation}
\Psi_{R}\equiv {\psi_1\choose {\psi_2}}_{R} = 
F_{R}={f_1\choose {f_2}}_{R}\, .\label{psr}
\end{equation}
With these fields we can now construct the most general gauge 
invariant chiral Lagrangian that includes the electroweak 
couplings of the top quark up to dimension four \cite{malkawi}.

\begin{eqnarray}
{\cal L}^{(4)}&=&i\overline{t}\gamma^{\mu}\left ( \partial_{\mu}
 +i\frac{2s_w^2}{3}{\cal A}_{\mu}\right) t
+i\ov {b}\gamma^{\mu}\left (\partial_{\mu}-i\frac{s_w^2}{3}
{\cal A}_{\mu}\right ) b\nonumber \\
&&-\frac{1}{2} \left (1-\frac{4s_w^2}{3}+{\kln}\right)
\ov {t_{L}} \gamma^{\mu} t_{L}{{\cal Z}_{\mu}}
 -\frac{1}{2}\left ( \frac{-4s_w^2}{3}+{\krn}\right ) \ov {{t}_{R}}
\gamma^{\mu} t_{R}{{\cal Z}_{\mu}} \nonumber \\
&&-\frac{1}{2}\left( -1+\frac{2s_w^2}{3}\right)
\ov {b_{L}}\gamma^{\mu} b_{L}{{\cal Z}_{\mu}}
-\frac{s_w^2}{3}\overline{b_{R}}\gamma^{\mu} b_{R}
{{\cal Z}_{\mu}}\nonumber \\
&&-\frac{1}{\sqrt{2}}\left (1+\kappa_{L}^{\rm {CC}}\right )
\ov {{t}_{L}} \gamma^{\mu} b_{L}
{{\cal W}_{\mu}^+}-\frac{1}{\sqrt{2}}
\left( 1+{\kappa_{L}^{\rm {CC}}}^{\dagger}\right)
\ov {{b}_{L}}\gamma^{\mu}t_{L}{{\cal W}_{\mu}^-} \nonumber \\
&&-\frac{1}{\sqrt{2}}\kappa_{R}^{\rm {CC}}
\ov {{t}_{R}}\gamma^{\mu} b_{R}
{{\cal W}_{\mu}^+}-\frac{1}{\sqrt{2}}
{\kappa_{R}^{\rm {CC}}}^{\dagger}\ov {{b}_{R}}
\gamma^{\mu} t_{R}{{\cal W}_{\mu}^-} \nonumber \\
&&-m_t \overline{t} t -m_b \overline{b} b \; . 
\label{eq2} 
\end{eqnarray}
In the above equation, the coefficients $\kappa_{L}^{\rm {NC}}$,
$\kappa_{R}^{\rm {NC}}$, $\kappa_{L}^{\rm {CC}}$,
and $\kappa_{R}^{\rm {CC}}$ parametrize possible deviations from
the SM predictions~\cite{malkawi}.

The constraints on the $\kappa$'s from the LEP/SLC data and from
the recent measurement on ${\rm Br}(b\ra s\gamma)$ \cite{recent},
are given in Refs. \cite{malkawi} and \cite{fuj},
respectively.   (We shall come back to these constraints later
in the section of Conclusions.)  As mentioned in the Introduction,
in this paper we are setting these  $\kappa$ 
coefficients to be zero in ${\cal L}^{(4)}$, which will be denoted as
${\cal L}^{(4)}_{SM}$ from now on.

In the next section we will construct the complete set of independent 
operators of dimension 5, such that the complete effective Lagrangian 
relevant to this work will be
\begin{equation}
{\cal L}_{eff}\; = \;{\cal L}^{B}\;+\; {\cal L}^{(4)}_{SM}\;+\;{\cal L}^{(5)}
\label{complang} \; ,
\end{equation}
where ${\cal L}^{(5)}$ denotes the higher dimensional operators.

\section{Dimension five operators}
\indent

Our next task is to find all the possible dimension five hermitian 
interactions that involve the top quark and the fields 
${\cal W}_{\mu}^{\pm}$,  ${\cal Z}_{\mu}$ and ${\cal A}_{\mu}$.  
Notice that the gauge transformations associated with these and the 
composite fermion fields ( Eq. ~(\ref{eq1}) ) are dictated simply by the
${\rm {U(1)}_{em}}$ group.    We will follow a procedure similar to the
one in Ref. \cite{buchmuller}, which consists of constructing all
possible interaction terms that satisfy the required gauge invariance,
 and that are not equivalent to each other. 
The criterion for equivalence is based on the equations of motion
(see Appendix A) and on partial integration.
As for the five dimensions in these operators,
three will come from the fermion fields, and the
other two will involve the gauge bosons.  To make a clear and
systematic characterization,  let us recognize the only three
possibilities for these two dimensions:

\noindent
(1) Operators with two boson fields.\\
(2) Operators with one boson field and one derivative.\\
(3) Operators with two derivatives.\\

\noindent
(1)  {\bf Two boson fields.}
\indent 
First of all, notice that the ${\cal A}_{\mu}$ field gauge transformation 
( Eq. ~(\ref{atransf}) ) 
will restrict the use of this field to covariant derivatives only.  
Therefore, except for the field strength term $\cal A_{\mu\nu}$ only the 
$\cal Z$ and $\cal W$ fields can appear multiplying the fermions in any 
type of operators.   
Also, the only possible Lorentz structures are given in terms of 
$g_{\mu \nu}$ and $\sigma_{\mu \nu}$  tensors.  We do not
need to consider the tensor product of $\gamma_{\mu}$'s since
\begin{equation} 
\aslash \bslash =  g_{\mu \nu} a^{\mu} b^{\nu} - i\sigma_{\mu \nu} 
a^{\mu} b^{\nu}\; .\label{idesig}
\end{equation}  
Finally,  we are left with only three possible combinations: \\
(1.1) two ${\cal Z}_{\mu}$'s, \\
(1.2) two ${\cal W}_{\mu}$'s,  and \\
(1.3) one of each.\\
Let us write down the corresponding operators for each case:

\indent
(1.1)  Since $\sigma_{\mu \nu}$ is antisymmetric, only the 
$g_{\mu \nu}$ part is non-zero:\footnote{ 
In the next section we will write explicitly the $h.c.$ parts.}
\begin{eqnarray}
O_{g{\cal Z}{\cal Z}} =
\bar t_L  t_R {\cal Z}_{\mu}{\cal Z}^{\mu} + h.c.
\label{ozz}
\end{eqnarray} 
\indent
(1.2) Here, the antisymmetric part is non-zero too:
\begin{eqnarray}
O_{g{\cal W}{\cal W}} &=& \bar t_L  t_R {\cal W}_{\mu}^{+} 
{\cal W}^{- \mu } + h.c. \label{opgww} \\
O_{\sigma {\cal W}{\cal W}} &=& 
\bar t_L  {\sigma}^{\mu\nu} t_R {\cal W}_{\mu}^{+} 
{\cal W}^{-}_{\nu } + h.c. \label{opsww}
\end{eqnarray} 
\indent
(1.3) In this case we have two different quark fields, therefore
we can distinguish two different combinations of chiralities:
\begin{eqnarray}
O_{g{\cal W}{\cal Z} L (R)} &=& 
\bar t_{L(R)}  b_{R(L)} {\cal W}_{\mu}^{+} {\cal Z}^{\mu } + h.c.
\label{ogwz} \\
O_{\sigma {\cal W}{\cal Z} L (R)} &=& 
\bar t_{L(R)} {\sigma}^{\mu\nu} b_{R(L)} {\cal W}^{+}_{\mu}
{\cal Z}_{\nu} + h.c. \label{oswz}
\end{eqnarray} 
\\
(2) {\bf One boson field and one derivative.}
\indent The obvious distinction arises:\\  
(2.1) the derivative acting on a fermion field, and\\
(2.2) the derivative acting on the boson.\\

\indent
(2.1) From Eqs.~(\ref{atransf}) and~(\ref{eq1}),
the covariant derivative for the fermions is given by
\footnote{To simplify notation we will use the same
symbol $D_{\mu}$ for all covariant derivatives. 
Identifying which derivative we are referring to 
should be straightforward, e.g. $D_{\mu}$ in Eq. (\ref{derf}) is
different from $D_{\mu}$ in Eq. (\ref{covderw}).}
\begin{eqnarray}
D_{\mu} f = (\partial_{\mu} + i Q_f s_w^2 {\cal A}_{\mu}) f
\nonumber ,\\
{\overline {D_{\mu} f}} = {\bar f}(\stackrel{\leftarrow}
{\partial}_{\mu} - iQ_f s_w^2 {\cal A}_{\mu}).\label{derf}
\end{eqnarray}
Notice that it depends on the fermion charge $Q_f$, 
hence the covariant derivative for the 
top quark is not the same as for the bottom quark; partial integration 
could not relate two operators involving derivatives on different quarks. 
Furthermore, by looking at the equations of motion we can immediately 
see, for example, that operators of the form $ \bar f {\zslash}{\Dslash} f $ 
or ${\bar f}^{(up)}{\wslash}^{+} {\Dslash} {f}^{(down)}$  are
equivalent to operators with two bosons, which have all been
considered already.  Following the latter statement and bearing in
mind the identity of Eq.~(\ref{idesig}) 
we can see that only one Lorentz structure needs to be
considered here, either one with $\sigma_{\mu \nu}$ or one
with $g_{\mu\nu}$.  Let us choose the latter. 
\begin{eqnarray}
O_{{\cal W}D b L (R)} &=&{\cal W}^{+ \mu} \bar t_{L(R)} 
D_{\mu} b_{R(L)}  + h.c.  \label{odbw} \\
O_{{\cal W}D t R (L)} &=&{\cal W}^{- \mu} \bar b_{L(R)} 
D_{\mu} t_{R(L)}  + h.c. \label{odtw} \\
O_{{\cal Z}D f } &=& {\cal Z}^{\mu} \bar t_{L} D_{\mu} t_{R} + h.c.
\label{odfz}
\end{eqnarray}
Of course, the $\cal A$ field did not appear.  
Remember that its gauge transformation prevents 
us from using it on anything that is not a covariant derivative or a 
field strength ${\cal A}_{\mu\nu}$. \\  
 
\indent
(2.2) Since $\cal W$  transforms as a field with electric charge one, 
the covariant derivative is simply given by (see Eq. ~(\ref{wtransf}) ):
\begin{eqnarray}
D_{\mu} {\cal W}_{\nu}^{+} &=& 
(\partial_{\mu} + i s_w^2 {\cal A}_{\mu}) {\cal W}_{\nu}^{+}
\nonumber \\
D^{\dagger}_{\mu} {\cal W}_{\nu}^{-} &=& (\partial_{\mu}
- is_w^2 {\cal A}_{\mu}) {\cal W}_{\nu}^{-}\label{Dw}
\end{eqnarray}

Obviously, since the neutral $\cal Z$ field is invariant under the $G$
group transformations [cf. Eq. ~(\ref{ztransf})], we could always
add it to our covariant derivative:
\begin{eqnarray}
D^{(\cal Z)}_{\mu} {\cal W}_{\nu}^{+} &=& 
(\partial_{\mu} + i s_w^2 {\cal A}_{\mu} + i a {\cal Z}_{\mu}) 
{\cal W}_{\nu}^{+} \, \nonumber 
\end{eqnarray}
where $a$ would stand for any complex constant.  
Actually, considering this second  
derivative would insure the generality of our analysis, since for
example by setting $a = c_w^2$ and comparing with
Eqs. ~(\ref{b1}) and ~(\ref{b2}) we would automatically include
the field strength term\footnote{From Eqs. ~(\ref{wdef})
and ~(\ref{wstrength}), we write ${\cal W}^{\pm}_{\mu\nu} = 
\frac{1}{\sqrt{2}}({\cal W}^1_{\mu\nu} \mp i {\cal W}^2_{\mu\nu})$.}
\begin{eqnarray}
{\cal W}_{\mu \nu}^{\pm} = 
{\partial}_{\mu} {\cal W}^{\pm}_{\nu} - 
{\partial}_{\nu} {\cal W}^{\pm}_{\mu} 
\pm i ( {\cal W}^{\pm}_{\mu} {\cal W}^{3}_{\nu} - 
{\cal W}^{3}_{\nu} {\cal W}^{\pm}_{\mu} ) = 
D^{(\cal Z)}_{\mu} {\cal W}_{\nu}^{\pm} - 
D^{(\cal Z)}_{\nu} {\cal W}_{\mu}^{\pm} .
\end{eqnarray} 
However, this extra term in the covariant derivative
would only be redundant.  
We can always decompose any given operator written in terms of  
$D_{\mu}^{(\cal Z)}$ into the sum of the same operator in terms of the 
original $D_{\mu}$ plus another operator of the form  
$O_{g{\cal W}{\cal Z} L (R)} $ or $O_{\sigma {\cal W}{\cal Z} L (R)}$ 
[cf. Eqs. ~(\ref{ogwz}) and ~(\ref{oswz})].  
Therefore, we only need to consider the covariant
derivative (\ref{Dw}) for the charged boson and still maintain the
generality of our characterization.
For the neutral ${\cal Z}$ boson,
the covariant derivative is just the ordinary one, 
\begin{eqnarray}
D_{\mu} {\cal Z}_{\nu} &=& \partial_{\mu} {\cal Z}_{\nu} .
\end{eqnarray}

The case for the ${\cal A}$ boson is nevertheless different. 
Being the field that makes possible the ${\rm {U(1)}_{em}}$ covariance
in the first place, it cannot be given any covariant derivative itself.  
For ${\cal A}$, we have the field strength: 
\begin{eqnarray}
{\cal A}_{\mu \nu} &=& 
{\partial}_{\mu}{\cal A}_{\nu} - {\partial}_{\nu}{\cal A}_{\mu}\; ,
\nonumber
\end{eqnarray}

Finally, we can now write the operators with the covariant 
derivative-on-boson terms.  
Unfortunately, no equations of motion can help us reduce the
number of independent operators in this case, and we have to
bring up both the $\sigma_{\mu\nu}$ and
the $g_{\mu\nu}$ Lorentz structures.
\begin{eqnarray}
O_{\sigma D {\cal Z}} &=& 
\bar t_L \sigma^{\mu \nu} t_R {\partial}_{\mu} {\cal Z}_{\nu} + h.c. \\
O_{g D {\cal Z}} &=& \bar t_L  t_R {\partial}_{\mu} {\cal Z}^{\mu} + h.c. \\
O_{\sigma D {\cal W} L(R)} &=& 
\bar t_{L(R)} \sigma^{\mu \nu} b_{R(L)} D_{\mu} {\cal W}^{+}_{\nu} + h.c. \\
O_{g D {\cal W}L(R)} &=& 
\bar t_{L(R)}  b_{R(L)} D_{\mu} {\cal W}^{+ \mu} +h.c. \\
O_{ {\cal A}} &=& \bar t_L \sigma^{\mu \nu} t_R  {\cal A}_{\mu \nu} + h.c.
\label{oa1}
\end{eqnarray}
\\
(3) {\bf Operators with two derivatives.}\\

As it turns out, all operators of this kind are equivalent to
the ones already given in the previous cases. 
Here, we shall present the argument of why 
this is so.  First of all, we only have two possibilities:\\
(3.1) one derivative acting on each fermion field,  and \\
(3.2) both derivatives acting on the same fermion field.\\

\indent
(3.1)  Just like in the  case (2.1) above,
we first notice that an operator 
of the form ${\bar f}\stackrel{\leftarrow}{\Dslash} {\Dslash} f$
can be decomposed into operators of the previous cases
(1.1), (1.2) and (1.3) by means of the equations of motion.
Therefore, we only have to consider one of two options, 
either ${\overline {D_{\mu} f}} \sigma^{\mu \nu} D_{\nu} f$, or   
${\overline {D_{\mu} f}} g^{\mu \nu} D_{\nu} f$. 
Let us choose the latter.   
By means of partial integration we can see that the term 
$(\partial_{\mu} \bar f) \partial^{\mu} f$ yields the same action
as the term $-\bar f \partial^{\mu} \partial_{\mu} f$,
and we only need to consider the 
case in which the covariant derivatives act on the same $f$,
which is just the type of operator to be considered next.\\

\indent
(3.2) Again, by using the equations of motion twice we can relate the 
operator  ${\bar f} {\Dslash} {\Dslash} f$ to operators of the type 
(1.1), (1.2) or (1.3).  
Either ${\bar f} \sigma^{\mu \nu} D_{\mu} D_{\nu} f$, 
or ${\bar f} D^{\mu} D_{\mu} f$ needs to be considered.  
This time we choose the former,
which can be proved to be nothing but the 
operator $O_{ {\cal A}}$ itself, i.e. Eq.~(\ref{oa1}).
\footnote{This can be easily checked by applying the
definition of $D_{\mu}$ in Eq.~(\ref{derf}).}.

\section{Hermiticity and CP invariance}
\indent

The  above list of the dimension 5 operators 
is complete in the sense that it includes 
all non-equivalent dimension five interaction terms
 that satisfy ${\rm{SU(2)}}_{L}\times {\rm{U(1)}}_{Y}$ gauge invariance. 
It is convenient now to analyze their CP properties.  
In order to make our study more systematic and clear we will re-write this 
list again, but this time we will display the added hermitian conjugate part 
in detail.  By doing this the CP transformation 
characteristics will be most clearly presented too.

Let us divide the list of operators in two: those with only the top quark, 
and those involving both top and bottom quarks.

\subsection {\bf Interactions with top quarks only}

\indent
Let's begin by considering the operator $O_{g {\cal Z} {\cal Z}}$.  
We will include an arbitrary constant coefficient $a$ which in principle 
could be complex, then
\begin{eqnarray}
O_{g{\cal Z}{\cal Z}} &{\sim}& 
a \bar t_L  t_R {\cal Z}_{\mu}{\cal Z}^{\mu} + 
a^{*} \bar t_R  t_L {\cal Z}_{\mu}{\cal Z}^{\mu} \nonumber \\
&=& Re(a) \bar t t  {\cal Z}_{\mu}{\cal Z}^{\mu}  + 
Im(a) i \bar t \gamma_5 t  {\cal Z}_{\mu}{\cal Z}^{\mu} \nonumber \, .
\end{eqnarray}
Our hermitian operator has naturally split into two
independent parts: 
one that preserves parity (scalar),
and one that does not (pseudoscalar).  
Also, the first part is CP even whereas the second one is odd.
The natural separation of these two parts happens to be a
common feature of all operators with only one type of
fermion field.  Nevertheless, not always will the parity
conserving part  also be CP even, as we shall soon see.

Below, the complete list of all 7 operators with only the top quark is given.  
In all cases the two independent terms are included; the first one is CP even, 
and the second one is CP odd. They are:
\begin{eqnarray}
O_{g{\cal Z}{\cal Z}} &=& {\frac {1}{\Lambda}}
Re(a_{zz1}) \bar t t {\cal Z}_{\mu}{\cal Z}^{\mu} \; +\; {\frac {1}{\Lambda}}
Im(a_{zz1}) i \bar t \gamma_5 t  {\cal Z}_{\mu}{\cal Z}^{\mu}\label{fir}, \\ 
O_{g{\cal W}{\cal W}} &=& {\frac {1}{\Lambda}}
Re(a_{ww1}) \bar t  t {\cal W}_{\mu}^{+} {\cal W}^{- \mu } \; + \; 
{\frac {1}{\Lambda}} 
Im(a_{ww1}) i \bar t \gamma_5 t {\cal W}_{\mu}^{+} {\cal W}^{- \mu }
\label{ogww} , \\
O_{\sigma {\cal W}{\cal W}} &=& {\frac {1}{\Lambda}}
Im(a_{ww2}) i \bar t  {\sigma}^{\mu\nu} t {\cal W}_{\mu}^{+} 
{\cal W}^{-}_{\nu }\; +\; {\frac {1}{\Lambda}} Re(a_{ww2}) \bar t  
{\sigma}^{\mu\nu} \gamma_5 t {\cal W}_{\mu}^{+} {\cal W}^{-}_{\nu } 
\label{osww}, \\ 
O_{{\cal Z}D f } &=&{\frac {1}{\Lambda}} Im(a_{z3}) i \bar t D_{\mu} t 
{\cal Z}^{\mu} \; + \; {\frac {1}{\Lambda}} Re(a_{z3}) \bar t \gamma_5
D_{\mu} t {\cal Z}^{\mu}  , \\
O_{g D {\cal Z}} &=& {\frac {1}{\Lambda}}
Im(a_{z4}) i \bar t \gamma_5 t {\partial}_{\mu} {\cal Z}^{\mu} \; +\; 
{\frac {1}{\Lambda}}Re(a_{z4}) \bar t  t {\partial}_{\mu} {\cal Z}^{\mu} , \\ 
O_{\sigma D {\cal Z}} &=& {\frac {1}{\Lambda}}
Re(a_{z2}) \bar t \sigma^{\mu \nu} t {\partial}_{\mu} {\cal Z}_{\nu} \; +\; 
{\frac {1}{\Lambda}} 
Im(a_{z2}) i \bar t \sigma^{\mu \nu} \gamma_5 t {\partial}_{\mu} 
{\cal Z}_{\nu} , \\ 
O_{{\cal A}} &=& {\frac {1}{\Lambda}}Re(a_{\cal A}) 
\bar t \sigma^{\mu \nu} t  {\cal A}_{\mu \nu}\; +\; {\frac {1}{\Lambda}} 
Im(a_{\cal A}) i \bar t \sigma^{\mu \nu}\gamma_5 t 
{\cal A}_{\mu \nu} \label{oa} .
\end{eqnarray}
Notice that in the operator $O_{g D {\cal Z}}$ the CP even part 
happens to be parity violating.  This is because under a CP 
transformation a scalar term $\bar t t$ remains intact, i.e. it does not 
change sign,  whereas a pseudoscalar term $\bar t \gamma_5 t$
changes sign.  
Also, the gauge bosons change sign under C, which is what makes the
scalar part of the $O_{g D {\cal Z}}$ operator to change sign under CP. 
On the contrary, 
the operator $O_{g{\cal Z}{\cal Z}}$ contains two bosons;
thus two changes of sign that counteract each other.  Therefore,
it is the scalar part of $O_{g{\cal Z}{\cal Z}}$ that is CP even.
In Table 1 we summarize in detail the discrete C, P and CP
symmetries of the above operators.

As for the size of these effective operators, based on the
naive dimensional analysis (NDA) their coefficients are of 
order $\frac{1}{\Lambda}$, where $\Lambda$ is the cut-off scale of the 
effective theory.  Therefore, the natural size of the normalized
coefficients (the $a$'s) is of order one.

\subsection {\bf Interactions with both top and bottom quarks}
 
\indent
Below, we show the next list of 12 operators with both top and bottom 
quarks.\footnote{
${\overline {D_{\mu} f}}_{R(L)}$ stands for 
$({D_{\mu} f_{R(L)}})^{\dagger}\gamma_0$; $\bar f_{R(L)}$ stands for 
$(f_{R(L)})^{\dagger}\gamma_0$.}
Again, we include an arbitrary complex coefficient $a$. They are: 
\begin{eqnarray}
O_{g{\cal W}{\cal Z} L (R)} &=& {\frac {1}{\Lambda}}
a_{wz1L(R)} \bar t_{L(R)}  b_{R(L)} {\cal W}_{\mu}^{+} {\cal Z}^{\mu } \; +\;
{\frac {1}{\Lambda}} 
a^{*}_{wz1L(R)} \bar b_{R(L)}  t_{L(R)} {\cal W}_{\mu}^{-} {\cal Z}^{\mu } ,\\
O_{\sigma {\cal W}{\cal Z} L (R)} &=& {\frac {1}{\Lambda}}a_{wz2L(R)} 
\bar t_{L(R)} {\sigma}^{\mu\nu} b_{R(L)} {\cal W}^{+}_{\mu}{\cal Z}_{\nu} + 
{\frac {1}{\Lambda}} a^{*}_{wz2L(R)} 
\bar b_{R(L)} {\sigma}^{\mu\nu} t_{L(R)} {\cal W}^{-}_{\mu}{\cal Z}_{\nu} , \\
O_{{\cal W}D b L (R)} &=& {\frac {1}{\Lambda}}
a_{bw3L(R)} {\cal W}^{+ \mu} \bar t_{L(R)} D_{\mu} b_{R(L)} \; +\; 
{\frac {1}{\Lambda}} 
a^{*}_{bw3L(R)} {\cal W}^{- \mu} {\overline {D_{\mu} b}}_{R(L)}  t_{L(R)}  , \\
O_{{\cal W}D t R (L)} &=& {\frac {1}{\Lambda}} a_{w3R(L)} 
{\cal W}^{- \mu} \bar b_{L(R)} D_{\mu} t_{R(L)} \; +\; {\frac {1}{\Lambda}}
a^{*}_{w3R(L)} {\cal W}^{+ \mu} {\overline {D_{\mu} t}}_{R(L)} b_{L(R)}  ,  \\
O_{\sigma D {\cal W} L(R)} &=& {\frac {1}{\Lambda}}a_{w2L(R)} 
\bar t_{L(R)} \sigma^{\mu \nu} b_{R(L)} D_{\mu} {\cal W}^{+}_{\nu} \; + 
{\frac {1}{\Lambda}} a^{*}_{w2L(R)} \bar b_{R(L)} \sigma^{\mu \nu} t_{L(R)} 
D^{\dagger}_{\mu} {\cal W}^{-}_{\nu}  ,\\
O_{g D {\cal W}L(R)} &=& {\frac {1}{\Lambda}} a_{w4L(R)} 
\bar t_{L(R)}  b_{R(L)} D_{\mu} {\cal W}^{+ \mu} \; +\;{\frac {1}{\Lambda}} 
a^{*}_{w4L(R)} \bar b_{R(L)}  t_{L(R)} D^{\dagger}_{\mu} {\cal W}^{- \mu}
\label{las} .  
\end{eqnarray}
In this case, if $a$ is real ($a = a^*$) then $O_{g{\cal W}{\cal Z} L (R)}$ 
and $O_{\sigma D {\cal W} L(R)}$ are both CP even, but 
$O_{\sigma {\cal W}{\cal Z} L (R)}$, $O_{{\cal W} D b L (R)}$, 
$O_{{\cal W} D t R (L)}$ and $O_{g D {\cal W}L(R)} $ are odd. 
Just the other way around if $a$ is purely imaginary [cf. Table 1].

The dimension five Lagrangian ${\cal L}^{(5)}$ is simply the sum
of all these 19 operators ~( Eqs.~(\ref{fir}) to ~(\ref{las}) ), i.e.
\begin{equation}
{\cal L}^{(5)} = \sum_{i=1,19} O_i \label{allops}
\end{equation}
  
For the purpose of this study; to estimate the possible effects on the 
production rates of top quarks in  high energy collisions, only the CP 
conserving parts, which give 
imaginary vertices (as the SM ones), are relevant.   This is because
the amplitude squared depends linearly on the CP even terms,
but only quadratically on the CP odd 
terms, and the {\it no-Higgs} SM  (${\cal L}^{(4)}_{SM}$) 
interactions\footnote{Since in the unitary gauge ${\cal L}^{(4)}_{SM}$
reproduces the SM without the physical Higgs boson, we will refer to it as 
the {\it no-Higgs} SM.} 
are CP even when ignoring the CP-violating
phase in the CKM quark mixing matrix.
     
However, this does not mean that we cannot probe the CP violating 
sector; as a matter of fact, later on in the section of Numerical Results 
we will show one observable that depends linearly on the CP odd 
coefficients.   From now on, the appropriate CP even part
(either real or imaginary) is assumed for each
coefficient .   To simplify notation we will use the same label; $a_{zz1}$
will stand for $Re(a_{zz1})$, $a_{wz2L(R)}$ will stand for
$Im( a_{wz2L(R)} )$, and so on.  The only exception will be $a_{\cal A}$,
whose real part is recognized as proportional to the magnetic moment
of the top quark, and will be denoted by $a_m$.
It is thus understood that all coefficients below are real numbers.

In conclusion, the dimension 5 Lagrangian consists of 19 independent
operators which are listed from Eq. (\ref{fir}) to Eq. (\ref{las}).   Their
eigenvalues under the C, P and CP transformations are conveniently
listed in Table \ref{cp}.  Operators with top and bottom quarks (right hand
side of the Table), which are given in terms of the chiral components,
are not eigenvectors of the C nor P transformations; therefore,
only the CP eigenvalues are given.

Since the top quark is heavy, its mass of the order of the weak scale, 
it is likely that it will interact strongly with the Goldstone bosons which 
are equivalent to the longitudinal weak gauge bosons in the high
energy regime.  
In the rest of this paper, we shall study how to probe these
anomalous couplings from the production of top quarks via
the $V_L V_L$ fusion process, where $V_L$ stands for the
longitudinally polarized $W^{\pm}$ or $Z$ bosons.

\begin{table}[htbp]
\begin{center}
\vskip -0.06in
\begin{tabular}{|l|c|c|c|c||c|c|c|c|}\hline \hline
\multicolumn{2}{|c|} {\mbox{Operator}} & C & P & 
CP & \multicolumn{2}{|c|} {\mbox{Operator}} &
CP \\  \hline \hline
$O_{g D {\cal Z}}$  & $i \bar t \gamma_5 t {\partial}_{\mu} {\cal Z}^{\mu}$
&  $-$ & $-$ & $+$  & $O_{g D {\cal W}L(R)} $  &
$i \bar t_{L(R)}  b_{R(L)} D_{\mu} {\cal W}^{+ \mu} \, +h.c.$ & $+$ \\
$\;$  & $ \bar t  t {\partial}_{\mu} {\cal Z}^{\mu}$
&  $-$ & $+$ & $-$  & $\;$  &
$\bar t_{L(R)}  b_{R(L)} D_{\mu} {\cal W}^{+ \mu} \, +h.c.$ & $-$ \\ \hline
$O_{\sigma D {\cal Z}}$  &
$\bar t \sigma^{\mu \nu} t {\partial}_{\mu} {\cal Z}_{\nu}$
&  $+$ & $+$ & $+$  & $O_{\sigma D {\cal W} L(R)}$  &
$\bar t_{L(R)} \sigma^{\mu \nu} b_{R(L)} D_{\mu} {\cal W}^{+}_{\nu}
 \, +h.c.$ & $+$ \\
$\;$  &
$i \bar t \sigma^{\mu \nu} \gamma_5 t {\partial}_{\mu} {\cal Z}_{\nu} $
&  $+$ & $-$ & $-$ & $\;$  &
$i \bar t_{L(R)} \sigma^{\mu \nu} b_{R(L)} D_{\mu} {\cal W}^{+}_{\nu}
 \, +h.c.$ & $-$ \\ \hline
$O_{{\cal Z}D f }$  & $i \bar t D_{\mu} t {\cal Z}^{\mu}$
&  $+$ & $+$ & $+$ & $O_{{\cal W}D t R (L)} $  &
$i {\cal W}^{- \mu} \bar b_{L(R)} D_{\mu} t_{R(L)} \, +h.c.$ & $+$ \\
$\;$  & $\bar t \gamma_5 D_{\mu} t {\cal Z}^{\mu} $
&  $+$ & $-$ & $-$ & $\;$  &
${\cal W}^{- \mu} \bar b_{L(R)} D_{\mu} t_{R(L)} \, +h.c.$ & $-$ \\ \hline
$O_{g{\cal Z}{\cal Z}}$  & $\bar t t {\cal Z}_{\mu}{\cal Z}^{\mu}$
&  $+$ & $+$ & $+$ 
& $O_{{\cal W}D b L (R)}$  &
$i {\cal W}^{+ \mu} \bar t_{L(R)} D_{\mu} b_{R(L)} \, +h.c.$ & $+$ \\
$\;$  & $i \bar t \gamma_5 t  {\cal Z}_{\mu}{\cal Z}^{\mu}$
&  $+$ & $-$ & $-$ 
& $\;$  &
${\cal W}^{+ \mu} \bar t_{L(R)} D_{\mu} b_{R(L)} \, +h.c.$ & $-$ \\ \hline
$O_{g{\cal W}{\cal W}}$  & $\bar t  t {\cal W}_{\mu}^{+} {\cal W}^{- \mu }$
&  $+$ & $+$ & $+$ 
& $O_{g{\cal W}{\cal Z} L (R)} $  &
$\bar t_{L(R)}  b_{R(L)}{\cal W}_{\mu}^{+} {\cal Z}^{\mu } \,+h.c.$ & $+$ \\
$\;$  & $i \bar t \gamma_5 t {\cal W}_{\mu}^{+} {\cal W}^{- \mu }$
&  $+$ & $-$ & $-$ 
& $\;$  &
$i \bar t_{L(R)}  b_{R(L)} {\cal W}_{\mu}^{+} {\cal Z}^{\mu }
\, +h.c.$ & $-$ \\ \hline
$O_{\sigma {\cal W}{\cal W}}$  &
$i \bar t  {\sigma}^{\mu\nu} t {\cal W}_{\mu}^{+} {\cal W}^{-}_{\nu }$
&  $+$ & $+$ & $+$ 
& $O_{\sigma {\cal W}{\cal Z} L (R)}$  &
$i \bar t_{L(R)} {\sigma}^{\mu\nu} b_{R(L)} {\cal W}^{+}_{\mu}
{\cal Z}_{\nu} \, +h.c.$ & $+$ \\
$\;$  & $\bar t  {\sigma}^{\mu\nu} \gamma_5 t
{\cal W}_{\mu}^{+} {\cal W}^{-}_{\nu }$ &  $+$ & $-$ & $-$ & $\;$  &
$\bar t_{L(R)} {\sigma}^{\mu\nu} b_{R(L)} {\cal W}^{+}_{\mu}
{\cal Z}_{\nu}  \, +h.c.$ & $-$ \\ \hline
$O_{{\cal A}}$  & $\bar t \sigma^{\mu \nu} t  {\cal A}_{\mu \nu}$
&  $+$ & $+$ & $+$ 
& $-$  &
$-$ & $\;$ \\
$\;$  & $i \bar t \sigma^{\mu \nu}\gamma_5 t {\cal A}_{\mu \nu} $
&  $+$ & $-$ & $-$ 
& $\;$  &
$\;$ & $\;$ \\ \hline
 \hline
\end{tabular}
\end{center}
\vskip 0.08in
\caption{The C, P and CP eigenvalues of all the dimension 5
operators.}
\label{cp}
\end{table}

\section{Probing the anomalous couplings}
\indent

\indent
In the following sections, we shall study the production rates of
$t\bar{t}$  ($t\bar{b}$ or $b\bar{t}$) from 
$W^{+}_L W^{-}_L$ or $Z_L Z_L$ ($W^{+}_L Z_L$ or
$W^{-}_L Z_L$) fusion processes in the TeV regime
for both the LHC and the LC.

\begin{figure}
\centerline{\hbox{
\psfig{figure=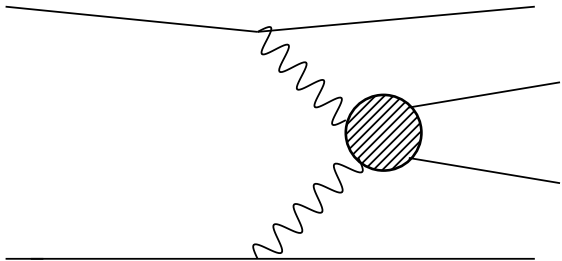,height=1.5in}}}
\caption{ Production of $t\bar{t}$  ($t\bar{b}$ or $b\bar{t}$) from 
the $V_L V_L$ fusion process.} 
\label{fusion}
\end{figure}

Before giving our analytical results (summarized in Appendices B and C),
we shall estimate the expected sizes of these tree level amplitudes
according to their high energy behavior.
A general power counting rule has been given that estimates the high
energy behavior of any amplitude $T$ \cite{poc} as:
\begin{eqnarray}
T=&& c_T v^{D_T}\displaystyle 
\left(\frac{v}{\Lambda}\right)^{N_{\cal O}}
\left(\frac{E}{v}\right)^{D_{E0}}
\left(\frac{E}{4 \pi v}\right)^{D_{EL}}
\left(\frac{M_W}{E}\right)^{e_v} H\left(\ln(\small {E/\mu})\right)
\label{eqcount}  \\ 
D_{E0}=&& 2+\sum_n {\cal V}_n
(d_n+\frac{1}{2}f_n-2)~, ~~~~ 
D_{EL}=2L~\nonumber,
\end{eqnarray}
where ${D_T}=4-e=0$ ($e$ is the number of external lines, 4 in our case),
${N_{\cal O}}= 0$  for all dimension 4 operators and  
${N_{\cal O}}= 1$  for all dimension 5 operators based upon the naive 
dimensional analysis (NDA) \cite{georgi},\footnote{
NDA counts $\Sigma$ as $\Lambda^0$, 
$D_\mu$ as $\frac{1}{\Lambda}$, and fermion fields as 
$\frac{1}{v\sqrt{\Lambda}}$. Hence, ${\cal W}^{\pm}$, 
${\cal Z}$ and ${\cal A}$ are also counted as 
$\frac{1}{\Lambda}$. After this counting,
one should multiply the result by 
$v^2 \Lambda^2$. Notice that up to the order of intent,
the kinetic term of the gauge boson fields and the mass
term of the fermion fields are two 
exceptions to the NDA, and are of order $\Lambda^0$.}
 $L=0$ is the number of loops in the diagrams,  
$ H\left(\ln(\small {E/\mu})\right) = 1$ 
comes from the loop terms (none in our case), ${e_v}$ accounts 
for any external $v_\mu$-lines (none in our case 
of $V_L V_L \ra t\ov t,\;t\ov b$),\footnote{
$v_\mu$ is equal to 
$\epsilon^{(0)}_{\mu}-\frac{k_\mu}{M_V}$, where $k_\mu$ is
the momentum of the gauge boson with mass $M_V$
and $\epsilon^{(0)}_{\mu}$ is its longitudinal
polarization vector.} 
${\cal V}_n$ is the number of vertices of type $n$ that contain 
$d_n$ derivatives and $f_n$ fermionic lines.  The dimensionless 
coefficient $~c_T~$ contains 
possible powers of gauge couplings ($g,g^\prime$) and Yukawa 
couplings ($y_f$) from the vertices 
of the amplitude $~T~$, which can be directly counted.

One important remark about the above formula is that it cannot be
directly applied to diagrams with external longitudinal $V_L$ lines. 
As explained in Ref.~\cite{poc}, a significant part of
the high energy behavior from diagrams with external $V_L$
lines is cancelled when
one adds all the relevant Feynman diagrams of the process;
this is just a consequence of the gauge symmetry of the
Lagrangian.   To correctly apply Eq.~(\ref{eqcount}), 
one has to make use of the Equivalence Theorem, and
write down the relevant diagrams with the corresponding
would-be Goldstone bosons.  Then, the true high energy
behavior will be given by the leading diagram.  (If there is
more than one leading diagram, there could be
additional cancellations).

\begin{figure}
\centerline{\hbox{
\psfig{figure=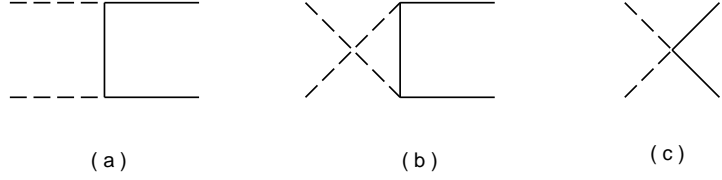,height=1in}}}
\caption{The corresponding Goldstone boson diagrams for 
$Z_L Z_L \rightarrow t\bar {t}$, i.e. $\phi^0\phi^0\ra t\ov t$.}
\label{power}
\end{figure}

Let us analyze the high energy behavior of the
$Z_L  Z_L  \ra t\bar {t}$ process in the context of the
dimension 4 couplings ${\cal L}^{(4)}$, as defined in Eq.~\ref{eq2}.
In Fig. \ref{power} we
show the corresponding Goldstone boson diagrams,
i.e. $\phi^{0} \phi^{0} \ra t\bar {t}$.  The $\phi^{0}$-$t$-$t$
vertex contains a derivative that comes from the
expansion of the composite fields [cf. Eq.~(\ref{zexp})],
and the associated $(d_n+\frac{1}{2}f_n-2)$ factor is
$d_n+\frac{1}{2}f_n-2 = 1+\frac{1}{2} 2-2 = 0$.  This means
$D_{E0}= 2$ for diagrams \ref{power}(a) and \ref{power}(b);
both grow as $E^2$ at high energies. 
The four point vertex for diagram \ref{power}(c) can come from
the mass term of the top quark or the
second order terms from the expansion of
${\cal Z}_{\mu}$ in the effective Lagrangian.
The $\phi^{0}$-$\phi^{0}$-$t$-${\ov t}$
vertex that comes from the mass
term $m_t {\ov {t}} t $ does not contain any derivatives,
hence the high energy behavior from this term goes like $E^1$.
The vertex that comes from the second order expansion
of a term like $\ov {t} \gamma^{\mu} t{{\cal Z}_{\mu}}$
contains one derivative, and the corresponding amplitude
\ref{power}(c) grows as $E^2$ in the high energy
region.   The conclusion is that
diagram \ref{power}(c) behaves like $E^2$ as well.
Seeing that there is more than one leading diagram we
can suspect that there may be additional cancellations.
How can we then obtain the correct high energy behavior
for the Goldstone boson scattering amplitudes?

To answer this question let us use an alternative non-linear
parametrization that is equivalent to ${\cal L}_{SM}^{(4)}$
(Eq. (\ref{eq2}) with the $\kappa$'s equal zero), in the sense that it
produces the exact same matrix elements \cite{chiral}, but with the
advantage that the couplings of the fermions with the
Goldstone bosons do not contain derivatives.  We can rewrite
the sum of ${\cal L}^{B}$ [cf.  Eq. (\ref{eq4})] 
and ${\cal L}_{SM}^{(4)}$ as:
\begin{eqnarray}
{\cal L}_{SM} \equiv {\cal L}_{SM}^{(4)} + {\cal L}^{B} =&&
{\ov {\Psi}_L} i \gamma^{\mu} D^L_{\mu} {{\Psi}_L} +
{\ov {\Psi}_R} i \gamma^{\mu} D^R_{\mu} {{\Psi}_R} - 
\left( {\ov {\Psi}_L} \Sigma M {{\Psi}_R} + h.c. \right) \nonumber \\
&&- {{1}\over {4}} W^a_{\mu\nu} W^{a\mu\nu} - {{1}\over {4}} B_{\mu\nu}
B^{\mu\nu} +  {{v^2}\over {4}} {\rm Tr}
\left( D_{\mu} \Sigma^{\dagger} D^{\mu} \Sigma \right)\;,
\label{noder} \\  M =&&
\left(
\begin{array}{r}
m_t \;\;\;\; 0\\
 0\;\;\;\; m_b
\end{array}
\right)\; ,\nonumber \\
D^L_{\mu} =&& \partial_{\mu} - i g  {\tau^a \over 2} W^a_{\mu}- 
i g^{\hspace{.5mm}\prime} {Y\over 2} B_{\mu}\; , \nonumber \\
D^R_{\mu} =&& \partial_{\mu} -i g^{\hspace{.5mm}\prime}Q_f B_{\mu}\; .
\nonumber 
\end{eqnarray}
Here, $Y=\frac{1}{3}$ is the hypercharge quantum number for the quark 
doublet, $Q_f$ is the electric charge of the fermion, $\Psi_L$ is the
linearly realized left handed quark doublet, and $\Psi_R$ is the
right handed singlet for top or bottom quarks
[cf. Eqs. (\ref{psi}) and (\ref{psr})].

As we shall see shortly, in the context of this Lagrangian there
is one (and only one) diagram with the leading high energy power.
Hence,  we do not expect any cancellations among diagrams and
it is possible to correctly predict the high energy behavior of the
scattering amplitude. Here is how it works:
When we expand the $\Sigma$ matrix field up to the second power 
[cf.  Eq.~(\ref{sigfield})] in the fermion mass term of Eq.~(\ref{noder}), 
we will notice two things: ({\it i}) the first power term gives to the 
vertex $\phi^{3}$-$t$-$t$, and associates the coefficient
$c_T\;=\; m_t / v$ to it; ({\it ii}) the second power term generates
the four-point vertex [cf. Fig. \ref{power}(c)] with a coefficient
$c_T\;=\; m_t^2 / v^2$ associated to it.
As it is well known, a $\ov {t} t \, = \, \ov {t}_R t_L  + \ov {t}_L t_R$
term always involves a chirality flip, therefore we readily recognize
that this four-point diagram will only participate when the chiralities
of the top and anti-top are different.  As $E \gg m_t$,
different chiralities imply equal helicities for the
fermion-antifermion pair.
Hence, for the case of opposite helicities we only count the power 
dependance for diagrams  2(a) and 2(b), and take the highest one.  For 
final state fermions of equal helicities we consider all three diagrams. 

The results are the following:  for diagrams  2(a) and 
2(b) we have $D_{E0}=2+(-1)+(-1)=0$ ,  
thus the amplitude $T_{\pm \mp}$ is of order $m_t^2/v^2$
(if there are no additional cancellations); which is 
the contribution given by the coefficients $c_T$ from both vertices.  
On the other hand, diagram  2(c) has $D_{E0}=2-1=1$;  
the equal helicities amplitude  $T_{\pm \pm}$ will be driven by this 
dominant diagram, therefore  $T_{\pm \pm} \; = \; m_t E/v^2$. 

For the other processes; $W^{+}_L W^{-}_L \rightarrow t \bar {t}$  and 
$W^{+}_L Z_L \rightarrow t \bar {b}$, the analysis is the same, except 
that there is an extra {\it s}-channel diagram [cf. Figs. \ref{fww}
and \ref{fwz}] whose high energy behavior is similar to the
diagrams \ref{power}(a) and \ref{power}(b).  Also, for the amplitude
of $W^{+}_L Z_L \rightarrow t \bar {b}$ no four-point
diagram \ref{power}(c) is generated; this means that its high
energy behavior can at most be of order $m_t^2/v^2$ as given
by diagrams \ref{power}(a) and \ref{power}(b).

In conclusion, in order to estimate the high energy behavior of the
$V_L  V_L  \ra t\bar {t}\, , \, t\bar {b}$ process, one has to write
down the relevant diagrams for 
$\phi \phi  \ra t\bar {t}\, , \, t\bar {b}$ and then
apply the power counting formula given in Eq. (\ref{eqcount}).
If more than one diagram have the same leading
power in $E$ then one can suspect possible additional
cancellations.  This is the case for the dimension 4 non-linear
chiral Lagrangian ${\cal L}^{(4)}_{SM}$ (Eq. (\ref{eq2}) with $\kappa$'s
equal to zero), for which all three diagrams
\ref{power}(a), (b) and (c) grow as $E^2$ at high energies.
Another gauge invariant Lagrangian for ${\cal L}^{(4)}_{SM} +
{\cal L}^{B}$ is given in Eq. (\ref{noder}) which gives the same
matrix elements for any physical process, but does not have
the problem of possible cancellations among the Goldstone boson
diagrams.  With this Lagrangian the power counting formula predicts a
leading $E^1$ behavior for $\phi^0 \phi^0 \ra t \bar t$
or $\phi^+ \phi^- \ra t \bar t$ (which originates from the 
four-point couplings that contributing
 to the diagram \ref{power}(c)), but only $E^0$ power for
 $\phi^\pm \phi^0 \ra t {\bar b} \, {\rm or} \, b {\bar t} $
(which does not have the diagram similar to \ref{power}(c)). 
This is verified in Appendix B.

Notice that, in general, if the dimension 4 anomalous
couplings $\kappa$'s are not zero, then there is no reason to expect
any cancellations among the Goldstone boson diagrams.
As a matter of fact, the calculated leading contributions from these
coefficients are of order $E^2$ and not $E^1$ 
[cf. Appendix B].\footnote{This is related
to the fact that non-zero anomalous $\kappa$ terms break the linearly
realized $SU(2)_L \times U(1)_Y$ gauge symmetry in
the interaction part of Eq. (\ref{noder}).
Notice that the $\kappa$ terms
respect this gauge symmetry only non-linearly.}

For the dimension 5 anomalous operators we do not
suspect {\it a priori} any cancellations
at high $E$ among Goldstone boson diagrams,
therefore we expect the parametrization used for our effective
operators to reflect the correct high energy behavior.  
Actually, the chiral Lagrangian parametrization given by 
Eq.~(\ref{complang}), which organizes the new
physics effects in the momentum expansion,
 is the only framework that allows the existence of such 
dimension 5 gauge invariant operators.   On the other hand,  
we know that as far as the {\it no-Higgs} SM contribution to these
{\it anomalous} amplitudes is concerned, the correct high energy
behavior is given by the equivalent  parametrization of
Eq.~(\ref{noder}).  We will therefore use the appropriate couplings
from ${\cal L}_{SM}$ and ${\cal L}^{(5)}$ in our next power counting
analysis.   Also, we are neglecting contributions of order
 $1/{\Lambda}^2$, which means that in diagrams
\ref{power}(a) and \ref{power}(b)  only one vertex is
anomalous.

Given one dimension 5 operator, it either involves two boson
fields (four-point operator), or one boson field and one derivative
(three-point operator).  Let us discuss four-point (4-pt) operators first.

There are three kinds of 4-pt operators: $O_{{\cal Z}{\cal Z}}$,
 $O_{{\cal W}{\cal W}}$ and  $O_{{\cal W}{\cal Z}}$.  Each of
them contributes to the $Z_L Z_L$, $W^{+}_L W^{-}_L$ and
$W^{+}_L Z_L$ fusion processes separately.  After expanding
the composite boson fields ${\cal Z}$ and ${\cal W}^{\pm}$
[cf. Eqs. (\ref{zexp}) and  (\ref{wexp})], we find that the
terms $\frac{4}{v^2} {\partial}_\mu{\phi}^{3}
{\partial}_\nu{\phi}^{3}$, $\frac{4}{v^2}
{\partial}_\mu{\phi}^{+} {\partial}_\nu{\phi}^{-}$ and
$\frac{4}{v^2} {\partial}_\mu{\phi}^{+}{\partial}_\nu{\phi}^{3}$
will contribute to a diagram of type \ref{power}(c) in each case.
Therefore, in the power counting formula (\ref{eqcount}), $d_n =2$,
$c_T = 4 a_O$ and $D_{E0}=2+(2+1-2)=3$, which means that
\begin{eqnarray}
T \sim {4 a_O}  
{{v}\over {\Lambda}} {\left( {E}\over {v}\right)}^3\; 
\label{fourpow} 
\end{eqnarray} 
for all these 4-pt operators.

Let us discuss the case of 3-pt operators by considering
one operator in particular:  $O_{{\cal Z}D f }$.  This
analysis will automatically apply to all the other six 3-pt; three with
the neutral ${\cal Z}_\mu$ boson, $O_{{\cal Z}D f }$,
$O_{g D {\cal Z}}$ and $O_{\sigma D {\cal Z}}$; and
three with the charged ${\cal W}_\mu$ boson,
$O_{{\cal W}D t}$, $O_{g D {\cal W}}$ and
$O_{\sigma D {\cal W}}$.
Using the expansions of the composite fields we obtain:
\begin{eqnarray}
a_{z3} i \bar t {\partial}_{\mu} t {\cal Z}^{\mu} \;\;\;\; =&&
 - \; \frac{g}{c_w} a_{z3} i{\ov {\psi}_t} {\partial}_{\mu}
{\psi}_t Z^{\mu} \; +\; \frac{2i}{v} a_{z3} \left[{\ov {\psi}_t}
{\partial}_{\mu} {\psi}_t \partial^{\mu} {\phi}^3 \right. 
\label{ogzexp} \\
 - {\ov {\psi}_t} {\gamma}_5 {\partial}_{\mu}
{\psi}_t {\phi}^3 \partial^{\mu} {\phi}^3\;\; -&&
\hspace{-0.5cm} \left.
 {\ov {\psi}_{tR}}{\partial}_{\mu}
{\psi}_{bL} {\phi}^{+} \partial^{\mu} {\phi}^3\; +\; 
{\ov {\psi}_t} {\partial}_{\mu}
{\psi}_t ({\phi}^{-} \partial^{\mu} {\phi}^{+} -
{\phi}^{+} \partial^{\mu} {\phi}^{-})\right] \;\; +
\cdot \cdot \cdot  \nonumber
\end{eqnarray}
Where $\psi_t$ ($\psi_b$) denotes the usual linearly realized top
(bottom) quark field.  There are more terms in Eq. (\ref{ogzexp}) that
participate in the Goldstone boson diagrams of interest, but
the ones shown are sufficient for our discussion. 
Notice that the first two terms on the right hand side of
Eq.  (\ref{ogzexp}) contribute to 3-pt vertices, the first one is for the
coupling of the top quark with the usual vector boson
field (the only non-zero term in the unitary gauge); the
second one represents the vertex of $O_{{\cal Z}D f }$
that enters in diagrams \ref{power}(a) and \ref{power}(b)
for ${\phi}^3 {\phi}^3 \ra t {\ov t}$, or in a
{\it u-channel} diagram like \ref{power}(b) for
${\phi}^{+} {\phi}^3 \ra t {\ov b}$.  The rest of the expansion
contains vertices with two or more boson fields.
In Eq.  (\ref{ogzexp}), we also show some of the 4-pt vertices
generated by $O_{{\cal Z}D f }$, which dominate the contribution
of this operator to the $V_L V_L$ fusion processes in the high
energy regime.
The last term,  ${\ov {\psi}_t} {\partial}_{\mu}
{\psi}_t ({\phi}^{-} \partial^{\mu} {\phi}^{+} -
{\phi}^{+} \partial^{\mu} {\phi}^{-})$,  comes from
the second order term in the expansion of ${\cal Z}_\mu$
[cf. Eq. (\ref{zexp})], and is responsible for the high
energy behavior of the {\it s-channel} diagram for
$W^{+}_L W^{-}_L \ra t {\ov t}$ [cf. Fig. \ref{fww}]. We
can infer that the other two 3-pt operators with the
${\cal Z}_\mu$ field can also contribute to all the $V_L V_L$
fusion processes.  However, because of the relation
${\epsilon}_{\mu} p^{\mu} = 0$ for the on-shell external
boson lines, the contributions of $O_{g D {\cal Z}}$ and
$O_{\sigma D {\cal Z}}$ vanish for $Z_L Z_L\ra t{\ov t}$
and  $W_L Z_L\ra t{\ov b}$. 

Notice that the expansion for ${\cal W}^{\pm}_\mu$ in
Eq. (\ref{wexp}) does not contain any term with $\phi^3$ alone;
hence, no operator with the field ${\cal W}^{\pm}_\mu$
can participate in the process $Z_L Z_L \ra t {\ov t}$
at tree level.  Except for this, the analysis on
$O_{{\cal Z}D f }$ applies equally to the operators with
${\cal W}^{\pm}_\mu$.   However, the contributions of
$O_{g D{\cal W}}$ and $O_{\sigma D{\cal W}}$ on the
process $W^{+}_L W^{-}_L \ra t {\ov t}$ vanish because
of the relation ${\epsilon}_{\mu} p^{\mu} = 0$ for the on-shell
external boson lines.

The analysis on the high energy behavior of the contributions
from $O_{{\cal Z}D f }$ to the scattering process
$Z_L Z_L \ra t {\ov t}$ is similar to the previous one for the
{\it no-Higgs} SM, in which we observed a distinction
between the  $T_{\pm \mp}$ and  $T_{\pm \pm}$ amplitudes.
The anomalous vertices generated by this operator contain
two derivatives, thus $(d_n+\frac{1}{2}f_n-2) = 1$.  
Then, $D_{E0}=2+1+(-1)=2$ for the first two diagrams
\ref{power}(a) and \ref{power}(b),
and $T_{\pm \mp}$ is of expected to be of order
\begin{eqnarray}
T_{\pm \mp} \sim {2 a_O}{{m_t}\over {v}}  
{{v}\over {\Lambda}} {\left( {E}\over {v}\right)}^2\; .\nonumber 
\end{eqnarray} 
On the other hand, diagram \ref{power}(c) comes from the
first 4-pt term in Eq. (\ref{ogzexp}).  Thus, we have
$(d_n+\frac{1}{2}f_n-2)=1$, $D_{E0}=2+1=3$, and the
predicted value for $T_{\pm \pm}$ is
\begin{eqnarray}
T_{\pm \pm} \sim {2 a_O}  
{{v}\over {\Lambda}} {\left( {E}\over {v}\right)}^3\; .\nonumber 
\end{eqnarray} 
Comparing with the estimate for 4-pt operators 
[cf. Eq. (\ref{fourpow})] we can observe that the only
difference is in the coefficient $c_T$ associated to them; for the
three-point operator (\ref{ogzexp}) $c_T = 2a_{O} $,
and for a four-point operator is twice as much.\footnote{
This difference in $c_T$ may be related to the fact that four-point
operators tend to give a bigger contribution to the helicity
amplitudes [cf. Eqs. (\ref{azz}) and (\ref{X}), for example].}

Other possible contributions that vanish have to do with the fact
that sometimes an amplitude can be zero from the product of
two different helicities of spinors.  For instance, by performing
the calculation of the amplitudes in the CM frame we
can easily verify that the spinor product
${\ov u}[\lambda =\pm1] v [\lambda =\mp1]$ 
vanishes for all $t {\ov t}$, $t {\ov b}$ and  $b {\ov t}$
processes.\footnote{${u}[\lambda = +1]$ denotes the spinor of a 
quark with right handed helicity.} 
This means that contributions from operators of the {\it scalar}-type,
like $O_{g{\cal W}{\cal Z} L (R)}$,
$O_{g{\cal Z}{\cal Z}}$,  $O_{g{\cal W}{\cal W}}$,  
$O_{{\cal Z}D f }$, and  
$O_{{\cal W}D t R (L)}$ will vanish for $T_{\pm\mp}$ amplitudes in
the {\it s}-channel and the four-point diagrams.

Furthermore, the relation ${\epsilon}_{\mu} p^{\mu} = 0$
applies to all the external on-shell boson lines; this makes the
contribution of operators with derivative on boson, such
as $O_{g D {\cal Z}} $ (our third case) and  $O_{g D {\cal W}L(R)}$,
to vanish in the {\it t}- and  {\it u}-channel diagrams. 
In principle, one would think that the exception could be
the {\it s}-channel diagram.  Actually, this is the case for the
operator $O_{g D {\cal W}L(R)}$ which contributes significantly to the
single top production process $W^{+}_L Z_L \rightarrow t {\bar {b}}$
via the s-channel diagram [cf. Table~\ref{opsderb}].  However, for the
 $O_{g D {\cal Z}} $ operator even this diagram vanishes; as can 
be easily verified by noting that the Lorentz contraction between the
boson propagator $-g_{\mu\nu}+k_{\mu} k_{\nu} / M_Z^2$ and the
tri-boson coupling is identically zero in the process
$W^{+}_L W^{-}_L \rightarrow t {\bar {t}}$.   
Therefore, for the $O_{g D {\cal Z}} $ operator all the 
possible diagrams vanish.

In Tables \ref{ops4}, \ref{opsderf}, and \ref{opsderb} we 
show the leading contributions (in powers of the CM energy $E$) 
of all the operators for each different process; those cells 
with a dash mean that no anomalous vertex generated by 
that operator intervenes in the given process, and those cells 
with a zero mean that the anomalous vertex intervenes in the 
process but the amplitude vanishes for any of the reasons 
explained above.   

\begin{table}[htbp]
\begin{center}
\vskip -0.06in
\begin{tabular}{|l||c||c||c||c||c||c|} \hline \hline
Process & ${\cal L}^{(4)}_{SM}$ & $O_{g{\cal Z}{\cal Z}}$
&$O_{g{\cal W}{\cal W}}$ &
$O_{\sigma {\cal W}{\cal W}}$ & $O_{g{\cal W}{\cal Z} L (R)} $ &
 $O_{\sigma {\cal W}{\cal Z} L (R)}$  \\ 
$\;$ & $\;$  & $a_{zz1}\times$ & $a_{ww1}\times$ &  $a_{ww2}\times$ &  
$a_{wz1L(R)}\times$ &  $a_{wz2L(R)}\times$ \\
 \hline\hline
$Z_L Z_L \rightarrow t \bar {t}$ & $m_t {E / {v^2}}$ & 
${E^3 / {v^2}{\Lambda}}$ & $-$& 
$-$ & $-$& $-$ \\  \hline
$W^{+}_L W^{-}_L \rightarrow t \bar {t}$  & $m_t {E / {v^2}}$ &$-$ & 
${E^3 / {v^2}{\Lambda}}$ & 
${E^3 / {v^2}{\Lambda}}$ & $-$ & $-$ \\ \hline
$W^{+}_L Z_L \rightarrow t\ov b$ & ${m_t^2 / {v^2}}$ &$-$ & $-$ & $-$ &
${E^3 / {v^2}{\Lambda}}$ & ${E^3 / {v^2}{\Lambda}}$ \\ \hline \hline
\end{tabular}
\end{center}
\vskip 0.08in
\caption{The leading high energy terms for the 4-point operators.}
\label{ops4}
\end{table}

\begin{table}[htbp]
\begin{center}
\vskip -0.06in
\begin{tabular}{|l||c||c||c||c|}\hline \hline
Process &   ${\cal L}^{(4)}_{SM}$ &
$O_{{\cal Z}D f } $ & $O_{{\cal W}D t R }$ &  $O_{{\cal W}D t L}$ \\ 
$\;$  & $\;$  &  $a_{z3}\times$ & $a_{w3R}\times$ & $a_{w3L}\times$ 
\\ \hline\hline
$Z_L Z_L \rightarrow t \bar {t}$ & $m_t {E / {v^2}}$ & 
${E^3 / {v^2}{\Lambda}}\;$& $-$ & $-$ \\ \hline
$W^{+}_L W^{-}_L \rightarrow t \bar {t}$ & $m_t {E / {v^2}}$ & 
${E^3 / {v^2}{\Lambda}}$ & ${E^3 / {v^2}{\Lambda}}\;$ & 
${{m_b E^2} / {v^2}{\Lambda}}\rightarrow 0$  \\ \hline 
$W^{+}_L Z_L \rightarrow t \ov b$ & ${m_t^2 /{v^2}}$ & 
${E^3 /{v^2}{\Lambda}}\;$ & 
${E^3 / {v^2}{\Lambda}}$ & ${E^3 / {v^2}{\Lambda}}$ \\ \hline\hline
\end{tabular}
\end{center}
\vskip 0.08in
\caption{The leading high energy terms for the operators 
with derivative-on-fermion.}
\label{opsderf}
\end{table}

\begin{table}[htbp]
\begin{center}
\vskip -0.06in
\begin{tabular}{|l||c||c||c||c||c||c|} \hline \hline
Process & ${\cal L}^{(4)}_{SM}$ &
$O_{g D {\cal Z}}$&$O_{g D {\cal W}L(R)}$&
$O_{\sigma D {\cal Z}}$&$O_{\sigma D {\cal W} L(R)} $&
$O_{ {\cal A}}$ \\
$\;$ & $\;$  & $a_{z4}\times$ & 
$a_{w4}\times$ &  $a_{z2}\times$ &  
$a_{w2}\times$ &  $a_{m}\times$ \\ \hline\hline
$Z_L Z_L\rightarrow t \bar {t}$ & $m_t {E / {v^2}}$ & 
$0$ & $-$& 
$0$ & $-$& $-$ \\  \hline
$W^{+}_L W^{-}_L \rightarrow t \bar {t}$  & $m_t {E / {v^2}}$ &$0$ & 
$0$ & ${E^3 / {v^2}{\Lambda}}$ & 
$0$ & ${E^3 / {v^2}{\Lambda}}$ \\ \hline
$W^{+}_L Z_L \rightarrow t \ov b$ & ${m_t^2  / {v^2}}$ &$0$ & 
$E^3/v^2{\Lambda}$ & $0$ & 
${E^3 / {v^2}{\Lambda}}$ & $-$ \\ \hline \hline
\end{tabular}
\end{center}
\vskip 0.08in
\caption{The leading high energy terms for the operators 
with derivative-on-boson.}
\label{opsderb}
\end{table}

In conclusion, based on the NDA \cite{georgi}
and the power counting rule \cite{poc}, we have found that the leading high 
energy behavior in the $V_L V_L \ra t\ov t\;or\;t\ov b$ scattering amplitudes 
from the {\it no-Higgs} SM operators (${\cal L}^{(4)}_{SM}$)
can only grow as $\frac{m_t E}{v^2}$ 
(for $T_{++}$ or $T_{--}$; $E$ is the CM energy of the top quark system), 
whereas the contribution from the dimension 5 operators (${\cal L}^{(5)}$) 
can grow as $\frac{E^3}{v^2\Lambda}$ in the high energy regime.   Let us 
compare the above results with those of the $V_LV_L \ra V_LV_L$ scattering 
processes.  For these $V_LV_L \ra V_LV_L$ amplitudes the leading
behavior at the lowest order gives $\frac{E^2}{v^2}$, and the contribution
from the next-to-leading order (NLO) bosonic operators gives
$\frac{E^2}{\Lambda^2}\frac{E^2}{v^2}$ \cite{poc}.   This indicates that
the NLO contribution is down by a factor of $\frac{E^2}{\Lambda^2}$ in
$V_LV_L \ra V_LV_L$.   On the other hand, the NLO fermionic contribution
in $V_L V_L \ra t\ov t\;or\;t\ov b$ is only down
 by a factor $\frac{E^2}{m_t \Lambda}$
which compared to $\frac{E^2}{\Lambda^2}$ turns
out to be bigger by a factor of $\frac{\Lambda}{m_t}\sim 4\sqrt{2}\pi$ for
$\Lambda \sim 4 \pi v$.\footnote{
For an energy $E$ of about $\Lambda /4$ or more this factor
$\frac{E^2}{m_t \Lambda} = \frac{M^{(5)}}{M^{(4)}}$ is actually
greater than one. $M^{(4)}$ and $M^{(5)}$ are the LO and NLO
amplitudes, respectively.}
   Hence, we expect that the NLO contributions in 
the $V_L V_L \ra t\ov t\;or\;t\ov b$ processes can be better measured 
(by about a factor of $10$) than the $V_LV_L \ra V_LV_L$ counterparts for 
some class of electroweak symmetry breaking models in which the NDA
gives reasonable estimates of the coefficients. 

As will be shown later, the coefficients of the NLO fermionic operators in 
${\cal L}^{(5)}$ can be determined via top quark production to an order 
of $10^{-2}$ or $10^{-1}$.   In contrast, the coefficients of the NLO
bosonic operators  are usually determined to about an order of $10^{-1}$  
or $1$ \cite{et,sss} via $V_LV_L \ra V_LV_L$ processes.  
Therefore, we conclude that the top quark production via longitudinal
gauge boson fusion $V_L V_L \ra t\ov t, t\ov b, \,{\rm or}\, b \ov t$
at high energy may be a better probe, for some classes of 
symmetry breaking mechanisms, than the scattering of longitudinal 
gauge bosons, i.e. $V_LV_L \ra V_LV_L$.

In the following section we shall study the production rates of $t\ov t$ 
pairs and single-$t$ or single-$\ov t$ events at future colliders like LHC 
and LC.   We will also estimate how precisely these NLO fermionic operators 
can be measured via the 
$V_L V_L \ra t\ov t, t\ov b, \,{\rm or}\, b \ov t$
 processes.  To reduce the number of 
independent parameters for our discussion, we shall assume an approximate 
custodial $SU(2)$ symmetry, so that the set of 19 independent coefficients 
will be reduced to 6 and given by  
$a_{zz1}, \, a_{z2}, \, a_{z3}, \, a_{z4}, \, a_{ww2}$, and   $a_m$.
However, for completeness we also 
give the leading high energy contributions of the helicity 
amplitudes for $Z_L Z_L \ra t\ov t$, $W^{-}_L W^{+}_L \ra t\ov t$,
$W^{+}_L Z_L \ra t\ov b$ and 
$W^{-}_L Z_L \ra b\ov t$ in Appendices B and C.  

\section{Underlying custodial symmetry}
\indent\indent

Here, we shall consider a special class of models of symmetry
breaking for which an approximate underlying custodial
symmetry \cite{dono} is assumed as suggested
by the low energy data \cite{malkawi}.

In addition to the gauge symmetry,   
the SM has an approximate global symmetry called 
the custodial symmetry which is responsible for the tree-level relation 
$\rho \simeq1$ [cf. Eq.~(\ref{yeq3})].  Actually, this symmetry is broken
by the hypercharge ($g^{\hspace{.5mm}\prime}$) and the mass splitting
($m_t \neq m_b$), but only slightly, so that $\rho$ remains about one 
at the one loop level.
After {\it turning off} the hypercharge coupling (i.e. set $s_w = 0$),
one can easily verify that
the global $SU(2)_L \times SU(2)_R$ symmetry is satisfied for the
dimension 4 Lagrangian ${\cal {L}}^{(4)}_{SM}$.\footnote{
To verify this, we
just need to use the transformation rules
$\Sigma \ra \Sigma^{'} \, =\, L \Sigma R^{\dagger}$ and
$F_L \ra F_L^{'} \, =\,R F_L$, where $L$ and $R$ are group elements 
of the global $SU(2)_L$ and $SU(2)_R$ symmetries, respectively.}
The fermion-gauge boson interactions are described by
\begin{equation}
{\cal L}^{(4custodial)}= \overline{F_L} \gamma^{\mu}
\left ( \; i\partial_{\mu}
\; - \frac{1}{2} {\cal W}_{\mu}^{a} \tau^a \; \right) 
F_L \, , \label{custodi}
\end{equation}
with the left handed doublet
\begin{equation}
F_{L}={f_1\choose {f_2}}_{L} \; \nonumber
\end{equation}
defined in Eq.~(\ref{psi}).

Notice that the only $SU(2)$ structure that satisfies the custodial symmetry
is the one given above.
If we want to introduce an anomalous interaction that satisfies this symmetry,
we must conform to this structure.  For example, let us consider the case
of the operators with derivative on boson $O_{g D {\cal Z}}$ and
$O_{g D {\cal W}L(R)}$, then we write:\footnote{
For the purpose of this
discussion we can replace $D_{\mu}$ by $\partial_{\mu}$.}
\begin{equation}
\kappa \overline{F_L}  g^{\mu\nu}
\partial_\mu {\cal W}_{\nu}^{a} \tau^a F_R \,+ h.c. \; = \;
\kappa \overline{F_L} g^{\mu \nu} \;  \left(
\begin{array}{r}
\partial_{\mu} {\cal W}_{\nu}^3 \;\;\;\; 
\sqrt{2} \partial_{\mu} {\cal W}_{\nu}^{+} \\
 \sqrt{2} \partial_{\mu} {\cal W}_{\nu}^{-} \;\;\;\; 
-\partial_{\mu} {\cal W}_{\nu}^3
\end{array}
\right)  \;  F_R \; +h.c. \; .  \nonumber
\end{equation}

As we can see, if we want our dimension 5 terms to obey this
global $SU(2)_L \times SU(2)_R$ symmetry, we have to introduce
the same anomalous interactions of the top quark to the much lighter
bottom quark.  Let us consider the case of
an underlying global $SU(2)_L \times SU(2)_R$
symmetry that is broken 
in such a way as to account for a negligible deviation
of the $\bbz$ vertex from its standard 
form.   Since the top quark acquires a
mass much heavier than the other quarks' masses, we expect the
new physics effects associated with the electroweak symmetry
breaking (EWSB) sector to be substantially greater for the
couplings (to the gauge bosons)
of this quark than for the couplings of 
all the others (including the bottom quark).
Therefore, it is probable that the underlying theory of 
particle physics respects the custodial symmetry, and the EWSB
mechanism introduces an effective interaction that  explicitly
breaks this symmetry in such a way as
to favor the deviation of the couplings
of the top quark more than the deviation 
of the other light quarks' couplings.

By adding the two possible breaking terms to this operator,\footnote{
Another term could be $\overline{F_L} g^{\mu\nu}\tau^3
\partial_{\mu} {\cal W}_{\nu}^{a} \tau^a \tau^3 F_R$, which contains two
symmetry breaking factors $\tau^3$ and will not be considered
in this work.} 
we obtain the effective dimension 5 Lagrangian as:
\begin{eqnarray}
{\cal L}^{(5deriv)}\;  =&&\;\kappa {\overline{F_L} } g^{\mu\nu}
\partial_\mu {\cal W}_{\nu}^{a} \tau^a F_R\, +\,
 \kappa_{1} {\overline{F_L}}  g^{\mu\nu} \tau^3
\partial_\mu {\cal W}_{\nu}^{a} \tau^a F_R \, + \,
\kappa_{2} {\overline{F_L}} g^{\mu\nu} \partial_\mu
{\cal W}_{\nu}^{a} \tau^a \tau^3 F_R \; \nonumber \\
+&& \; h.c. \\
=&& {\overline{F_L}} g^{\mu \nu} \;  \left(
\begin{array}{r}
(\kappa + \kappa_1 + \kappa_2 )
\partial_{\mu} {\cal W}_{\nu}^3 \;\;\;\; 
\sqrt{2} (\kappa + \kappa_1 - \kappa_2 )
\partial_{\mu} {\cal W}_{\nu}^{+} \\
 \sqrt{2} (\kappa - \kappa_1 + \kappa_2 )
\partial_{\mu} {\cal W}_{\nu}^{-} \;\;\;\; 
(-\kappa + \kappa_1 + \kappa_2 )
\partial_{\mu} {\cal W}_{\nu}^3
\end{array}
\right)  \;  F_R \; +h.c. \;,  \nonumber
\end{eqnarray}
where, in order to obtain a vanishing $\bbz$ coupling,
we require
\begin{equation}
\kappa \; = \; \kappa_1 \, +\, \kappa_2 \,. \nonumber
\end{equation}

Also, to simplify the discussion we assume
$\kappa_1 = \kappa_2$, and the conclusion is that in order to
keep the couplings $\bbz$ unaltered we have to impose the
condition
\begin{equation}
a_{z(2,3,4)} = \sqrt{2}  a_{w(2,3,4)L(R)}\,  
\end{equation}
to all the operators with derivatives.

The case for 4-point operators (contact terms) is somewhat different.  The
custodial Lagrangian in this case is of the form:
\begin{eqnarray}
{\cal L}^{(5custod)}=&&\kappa^{4pt.}_{1g} \overline{F_L} g^{\mu\nu}
{\cal W}_{\mu}^{a} \tau^a {\cal W}_{\nu}^{b} \tau^b F_R \; + \; 
\kappa^{4pt.}_{1\sigma} \overline{F_L}
\sigma^{\mu\nu} {\cal W}_{\mu}^{a} \tau^a {\cal W}_{\nu}^{b} \tau^b F_R
\nonumber \\
=&& \kappa^{4pt.}_{1g} \overline{F_L}
\left(
\begin{array}{r}
{\cal W}_{\mu}^{3} {\cal W}^{\mu 3} +
2 {\cal W}_{\mu}^{+} {\cal W}^{\mu -} \;\;\;\;\;\;\;\;\;\;\;\; 0\\
 0 \;\;\;\;\;\;\;\;\;\;\;\; {\cal W}_{\mu}^{3} {\cal W}^{\mu 3} +
2 {\cal W}_{\mu}^{+} {\cal W}^{\mu -}
\end{array}
\right) 
F_R  \nonumber \\
&& \qquad\qquad 
  + \kappa^{4pt.}_{1\sigma} \overline{F_L} \sigma^{\mu\nu}
\left(
\begin{array}{r}
2 \sigma^{\mu\nu} {\cal W}_{\mu}^{+} 
{\cal W}_{\nu}^{-} \;\;\;\;\;\;\;\;\;\;\; 0\\
 0 \;\;\;\;\;\;\;\;\;\;\;2 \sigma^{\mu\nu} 
{\cal W}_{\mu}^{+} {\cal W}_{\nu}^{-}
\end{array}
\right) 
F_R \, ,
 \label{custod5}
\end{eqnarray}
and for the breaking terms we can consider:
\begin{eqnarray}
{\cal L}^{(5contact)}=&& \sum_{c=g,\sigma}
c^{\mu\nu} \,\left( \;\; \kappa^{4pt.}_{2c} \overline{F_R} \tau^3
{\cal W}_{\mu}^{a} \tau^a {\cal W}_{\nu}^{b} \tau^b F_L\,+\,
\kappa^{4pt.\dagger}_{2c} \overline{F_L} {\cal W}_{\mu}^{a} \tau^a
{\cal W}_{\nu}^{b} \tau^b \tau^3 F_R \right. \nonumber \\
&& \qquad\qquad 
+ \left. \kappa^{4pt.}_{3c} \overline{F} {\cal W}_{\mu}^{a} 
\tau^a \tau^3
{\cal W}_{\nu}^{b} \tau^b F \; \; \right) \; ,
\end{eqnarray}
where $\kappa^{4pt.}_{2c}$ is complex and $\kappa^{4pt.}_{3c}$ is real.
As it turns out, in order to set the anomalous couplings of the bottom
quark equal to zero, we have to choose $\kappa^{4pt.}_{3c}=0$, and
$\kappa^{4pt.}_{2c}$ real and half the size of $\kappa^{4pt.}_{1c}$
(i.e. $\kappa^{4pt.}_{1c}=2 \kappa^{4pt.}_{2c}$ for $c=g,\sigma$). The
non-standard 4-point dimension 5 interactions will then have the
structure
\begin{equation}
\left(
\begin{array}{r}
c^{\mu\nu} {\cal W}_{\mu}^{3} {\cal W}_{\nu}^{3} +
2 c^{\mu\nu} {\cal W}_{\mu}^{+} {\cal W}_{\nu}^{-} 
\;\;\;\;\; 0\\
 0 \;\;\;\;\;\;\;\;\;\;\;\;\;\;\;\;\;\;\;\;\;\;\;\; 0
\end{array}
\right) 
\end{equation}
where $c^{\mu\nu}$ is either $g^{\mu\nu}$ or $\sigma^{\mu\nu}$.
This structure suggests $2 a_{zz1}  =  a_{ww1}$, and
$a_{wz1L(R)} = a_{wz2L(R)} = 0$ for $c^{\mu\nu}$ equal to $g^{\mu\nu}$.
For  $c^{\mu\nu}$ equal to $\sigma^{\mu\nu}$, it suggests that
$a_{ww2}$ can be of any value.

In conclusion, by assuming the dimension 5 operators are
the result of an underlying custodial symmetric theory that is broken
in such a way that at tree level
the $Z$-$b$-$b$ coupling does not get modified from its SM values,
we derive the following relations among the coefficients of these
anomalous couplings. They are:
\begin{eqnarray}
a_{z(2,3,4)} =&& \sqrt{2}  a_{w(2,3,4)L(R)}\; , \nonumber \\
2 a_{zz1}  =&& \, a_{ww1}\; , \label{custrel} \\
a_{wz1L(R)} =&& a_{wz2L(R)} = 0\; . \nonumber
\end{eqnarray}
After including the hypercharge interactions, we can see that
the set of independent coefficients has reduced from a
total of $19$ down to $6$ only.  These coefficients are
$a_{z(2,3,4)}$, $a_{zz1}$,$a_{ww2}$ and
$a_{m}$ (for the operator $O_{{\cal A}}$).

\section{ Amplitudes for $Z_L Z_L$, $W_L W_L$,
and $W_L Z_L$ fusion processes}
\indent\indent

Below, we present the helicity amplitudes for each $V_L V_L$ fusion
process.  We shall only consider the leading contributions in powers 
of $E$, the CM energy of the $V_L V_L$ system, coming
from both the {\it no-Higgs} SM (i.e. ${\cal {L}}^{(4)}_{SM}$) and
the dimension 5 operators.  For the latter, we assume an
approximate $\rm{SU(2)}$ custodial symmetry, as discussed in the
previous section,  so that only 6 independent coefficients
are relevant to our discussion. For completeness, in
Appendices B and C we provide the helicity amplitudes for the
general case (without assuming a custodial symmetry).

\subsection{  $Z_L Z_L \rightarrow t\bar{t}$}
\indent\indent

\begin{figure}
\centerline{\hbox{
\psfig{figure=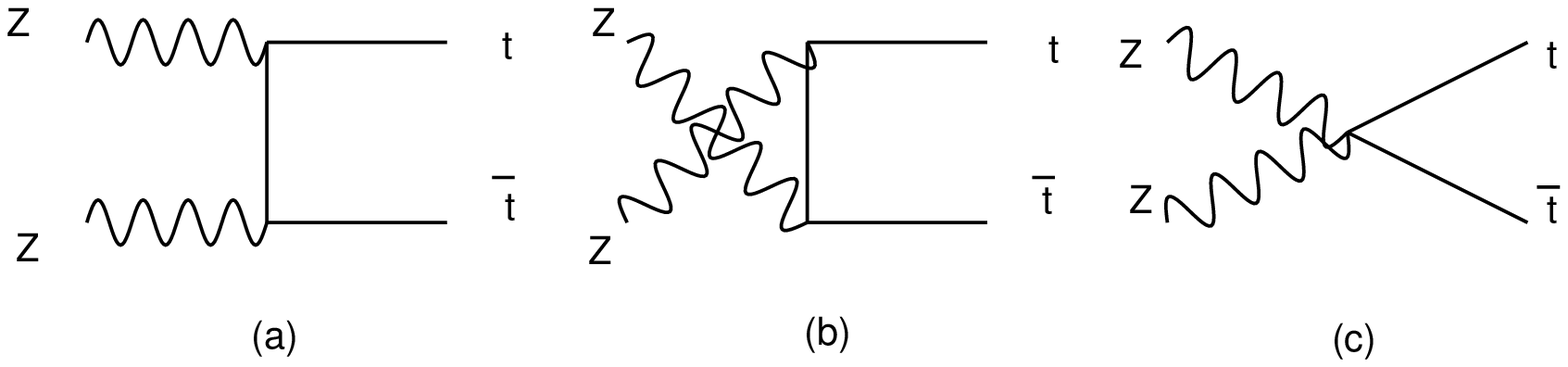,height=1in}}}
\caption{Diagrams for the $Z Z \rightarrow t\bar{t}$ process.}
\label{fzz}
\end{figure}

Fig. \ref{fzz} shows the diagrams associated to this process.  The
total amplitude $T$ is the sum of the ${\cal {L}}^{(4)}_{SM}$
contribution (denoted by $zz$), and the ${\cal {L}}^{(5)}$
contribution (denoted by $azz$).  In diagrams with
two vertices, only one anomalous vertex is considered at a time,
i.e. we neglect contributions suppressed by $1/{\Lambda}^2$.
We denote the helicity amplitudes by the helicities of the outgoing
fermions: the first (second)  symbol ($+$ or $-$) refers to the fermion on
top (bottom) part of the diagram.  A right handed fermion is labelled
by '$+$', and a left handed fermion by '$-$'.  For instance,
\begin{equation}
T_{zz++} = zz_{++} \, + \, azz_{++} \; ,
\end{equation}
where $zz_{++}$ is the ${\cal {L}}^{(4)}_{SM}$ contribution, and
$azz_{++}$ is the {\it anomalous} contribution to the helicity
amplitude $T (Z_L Z_L \ra t_{Right-handed} \bar{t}_{Right-handed} )$.
The same notation is used for the other two processes.

The leading contributions to the $Z_L Z_L \ra t \bar{t}$ helicity
amplitudes are:
\begin{eqnarray}
T_{zz++} =&& -T_{zz--} \;\;\;=\;\; {{{\it m_t}\,E}\over {{v^2}}} -
{{2 E^3}\over {{v^2}}} {X\over {\Lambda}}\; 
 , \nonumber \\
T_{zz+-} =&& T_{zz-+} \;\;\;=\;\; {{2\,\,{{m^2_t}}\,
{c_{\theta}} {s_{\theta}}}\over
{\left( {{4{{{c^2_{\theta}}}}\,m^2_t}\over {{E^2}}} + 
   {{{s^2_{\theta}}}} \right) \,{v^2}}} \; +0 \; ,
\label{azz}
\end{eqnarray}
where
\begin{equation}
X = {\it a_{zz1}}+\left({1\over 2}-{4\over 3} 
s^2_w\right){\it a_{z3}}\; ,
\label{X}
\end{equation}
and $E\,=\,\sqrt{s}$ is the CM energy of the $V_L V_L$
system.

Comparing with the results for $W^{+}_L W^{-}_L$
and  $W^{+}_L Z_L$ fusions, this is the amplitude that takes the
simplest form with no angular dependance.  Also, for this process
the assumption of an underlying custodial symmetry does not make
the anomalous contribution any different from the most general
expression given in Appendix C. 
This means that  new physics effects coming through
this process can only modify the S-partial wave amplitude (at the
leading order of $E^3$). 
Notice that at this point it is impossible to distinguish the effect of the
coefficient $a_{zz1}$ from the effect of the coefficient $a_{z3}$. 
However, in the next section we will show how to combine this
information with the results of the other processes, and obtain bounds
for each coefficient.  The reason why $azz_{\pm \mp}$ appear as
zero is explained in Appendix C.

\subsection{ \bf $W^{+}_L W^{-}_L \rightarrow t\bar{t}$}

\begin{figure}
\centerline{\hbox{
\psfig{figure=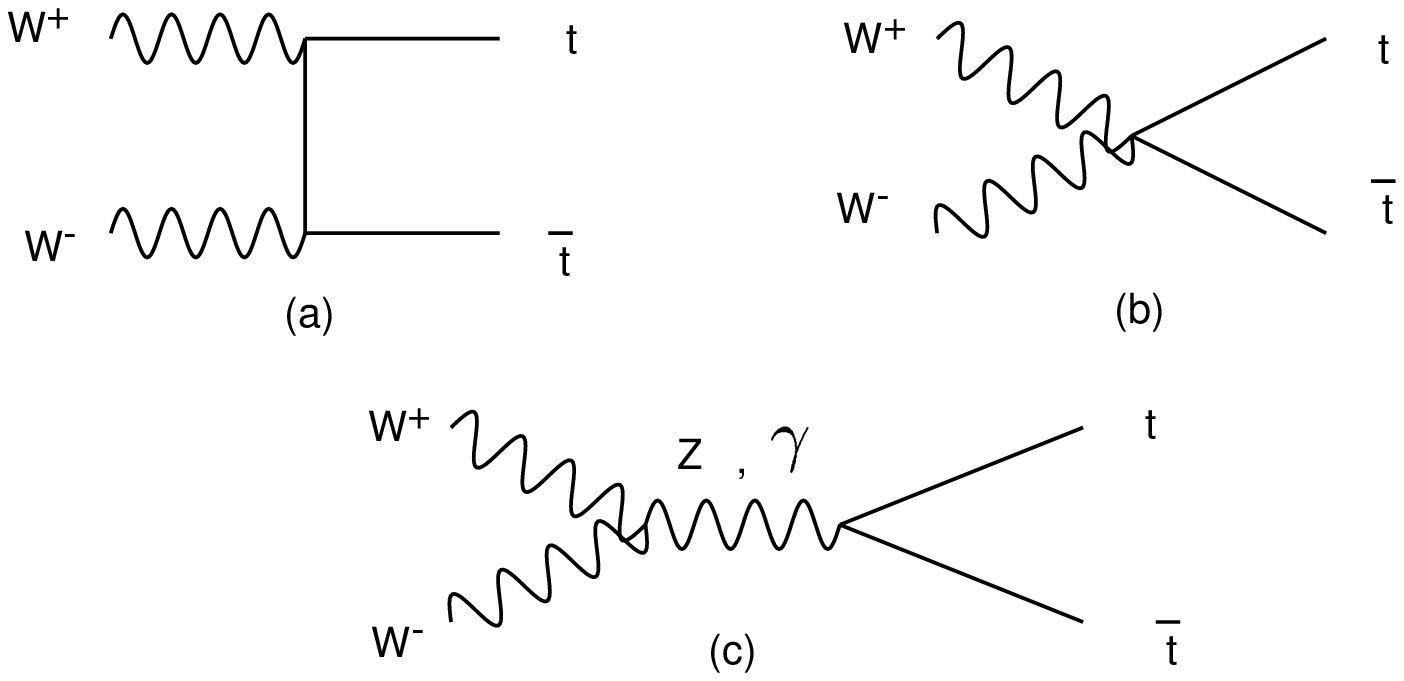,height=2in}}}
\caption{Diagrams for the $W W \rightarrow t\bar{t}$ process.}
\label{fww}
\end{figure}

\indent
The amplitudes of this process are similar to the ones of the previous
process except for  the presence of two {\it s}-channel diagrams
(see Fig. \ref{fww}), whose off-shell $\gamma$ and $Z$  propagators
allow for the additional contribution from the magnetic moment of the
top quark (${\it a_{m}}$) and the operator with derivative on boson
$O_{\sigma D {\cal Z}}$ (${\it a_{z2}}$).
Since these two operators are not of the $scalar$-type, 
we have a non-zero contribution to the $T_{\pm\mp}$ amplitudes.
Throughout this paper, the angle of scattering $\theta$ in all
processes is defined to be the one subtended between the
incoming gauge boson that appears on the top-left part of the
Feynman diagram ($W^{+}$ in this case) and the momentum of
the outgoing fermion appearing on the top-right part of the same
diagram ($t$ in this case); all in the CM frame of the
$V_L V_L$ pair.

The leading contributions to the various helicity amplitudes for this 
process are:
\begin{eqnarray}
T_{ww++} =&& -T_{ww--} \;\;\;=\;\; {{{m_t}\,E}\over {{v^2}}}\, -\,
{{4 \,{{E}^3}}\over {v^2}} 
{{\left(X_1 + X_m c_{\theta}\right)}\over {\Lambda}}\; ,
\nonumber \\
T_{ww+-} =&&{{2\,{{{m^2_t}}}\,{s_{\theta}}}\over 
{\left( {2{{{{m_b}}^2}}\over {{E^2}}} +\left(1 - {c_{\theta}} \right) \,
 \left( 1 - {2{{{m^2_t}}}\over {{E^2}}} \right)  \right)\, {v^2}}} \, + \,
 {{{8{E}^2} \over {v^2}} m_t s_{\theta}\,
{{\left( X_m - \frac{1}{4}{\it a_{z3}} \right)}\over{\Lambda}}}\; ,
\nonumber\\
T_{ww-+} =&& 0 \, + \, {{8{E}^2} \over {v^2}} m_t s_{\theta} 
{{X_m-\frac{1}{8}{\it a_{z3}}}\over {\Lambda}} \; ,
\end{eqnarray}
where $s_{\theta} = \sin{\theta}$, $c_{\theta} = \cos{\theta}$, and
\begin{eqnarray}
X_1=&& {\it  a_{zz1}}+ \frac{1}{8}{\it a_{z3}} \; ,\nonumber\\
X_m=&& {\it a_{m}} - \frac{1}{2}{\it a_{z2}}
+ \frac{1}{8}{\it a_{z3}}+ \frac{1}{2}{\it a_{ww2}}\; .
\label{Xprime}
\end{eqnarray}

Notice that the angular distribution of the leading contributions in the 
$T_{\pm\pm}$ amplitudes consists of the flat
component (S-wave) and the $d^1_{0,0}=\cos {\theta}$
component (P-wave).   The $T_{\pm\mp}$  
helicity amplitudes only contain the 
$d^1_{0,\pm 1}=-\frac{\sin {\theta}}{\sqrt{2}}$ component.  
This is so because the initial state consists of longitudinal gauge bosons 
and has zero helicity. 
The final state is a fermion pair so that the helicity of this state can be 
$-1$,  $0$, or $+1$.   Therefore, in high energy scatterings, the
anomalous dimension 5 operators only modify
(at the leading orders $E^3$ and $E^2$ ) the S- and
P-partial waves of the scattering amplitudes.   We also note that, as
expected from the discussion in section 5,
$aww_{\pm\pm}$ has an $E^3$
leading behavior, whereas $aww_{\pm\mp}$ goes like $E^2$.
Furthermore, the ${\cal {L}}^{(4)}_{SM}$ amplitudes are of
order $m_t E/v^2$ for $ww_{\pm\pm}$,
and $m_t^2/v^2$ for $ww_{+-}$.
($ww_{-+}$ is proportional to $m_b^2/v^2$ and is taken as zero.)
To calculate the event rate, we need to sum over
four helicity amplitudes squared, and
$\mid T_{\pm \pm , \pm \mp} {\mid}^2 = ww^2_{\pm \pm ,\pm \mp}
+2 ww_{\pm \pm , \pm \mp}\, aww_{\pm \pm , \pm \mp} +
O(1/ {\Lambda}^2)$.  Because $\mid ww_{ \pm \mp}\,
aww_{ \pm \mp}{\mid} \sim \frac{m_t^2}{E^2} \mid
ww_{ \pm \pm}\, aww_{ \pm \pm} {\mid}$, the
amplitude squared $\mid T_{\pm\mp} {\mid}^2$ is only
a few percent of the value of $\mid T_{\pm\pm} {\mid}^2$ for
$E \sim 1$ TeV. Thus, $\mid T_{\pm\mp} {\mid}^2$ will not
contribute largely to the total event rate, provided the
coefficients of the dimension 5 operators are of order one.

\subsection{ \bf $W^{+}_L Z_L \rightarrow t\bar{b}$}

\begin{figure}
\centerline{\hbox{
\psfig{figure=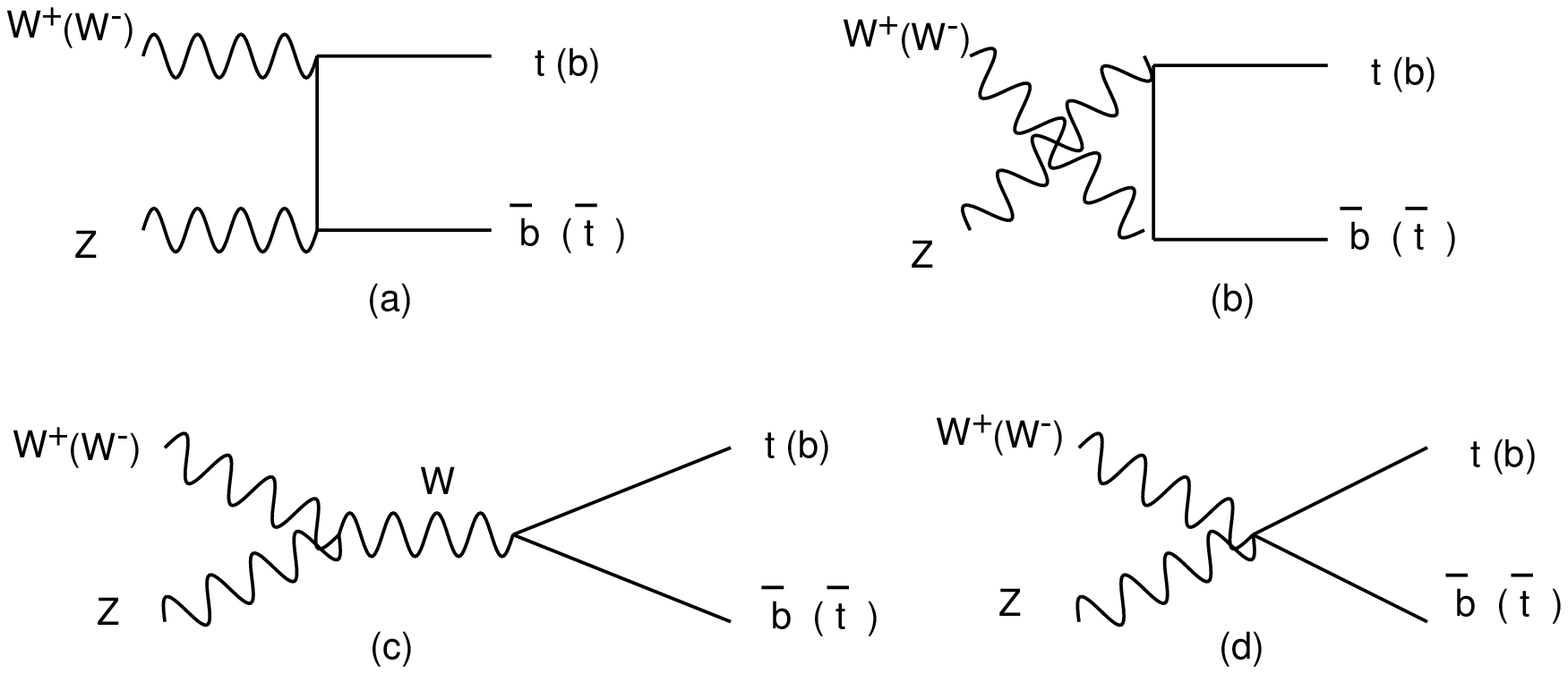,height=2in}}}
\caption{Diagrams for the $W Z \rightarrow t\bar{b}$ process.}
\label{fwz}
\end{figure}

\indent
Finally, we have the amplitudes for the
single-top quark production process 
$W^{+}Z \rightarrow t\bar{b}$ (which are just the
same as for the conjugate 
process $W^{-}Z \rightarrow b\bar{t}$). 
Fig. \ref{fwz} shows the diagrams
that participate in this process.

The leading contributions to the various helicity amplitudes for this 
are:\footnote{As shown in Eq. (\ref{custrel}), for models with this
approximate custodial symmetry, $a_{wz1L(R)} = a_{wz2L(R)} =0$,
so that the 4-point vertex diagram of Fig. \ref{fwz}(d) gives no
contribution.}
\begin{eqnarray}
T_{wzt++} =&&  -{{{\sqrt{2}}\,{m^3_t} \,
\left( 1 - {c_{\theta}} \right) }\over 
{E\,{{\left( 1 - {2{{{{m^2_t}}}}\over {{E^2}}} \right) }}\,
\left( 1 + {c_{\theta}} + {2{{{{m^2_t}}}}\over 
  {{E^2}}} \right) \,{v^2}}} \, -\,
{ \sqrt{2} {E}^3 \over {4 v^2}} 
{{\left(  X_2 + c_{\theta}  X_3  \right)}
\over {\Lambda}}\; ,\nonumber\\
T_{wzt--} =&& 0 \, -\, { \sqrt{2} {E}^3 \over {4 v^2}} 
{{\left( (4 s^2_w a_{z4} +\frac{2}{3} s^2_w - 1) +
(a_{z3} - 4 c^2_w a_{z2}) c_{\theta} \right)}\over {\Lambda}}\; ,
\nonumber\\
T_{wzt+-} =&& 0 \, +\, {{\sqrt{2} {\it E}^2}\over {4 v^2}} m_t s_{\theta}\,
{{\left( a_{z3}+ 4 c^2_w  a_{z2} \right)}\over {\Lambda}} \; ,
\nonumber\\
T_{wzt-+} =&&  -{{{\sqrt{2}}\,{{{m^2_t}}}\,{s_{\theta}}}\over 
  {{{\left( 1 - {2{{{{m^2_t}}}}\over {{E^2}}} \right) }}\,
\left(1 + {c_{\theta}} + {2{{{{m^2_t}}}}\over{{E^2} }} \right) v^2}} \, -\,
{{3 \sqrt{2} {E}^2}\over {4 v^2}} m_t s_{\theta}\,
{{ X_4 }\over {\Lambda}}\; ,
\label{awz}
\end{eqnarray}
where
\begin{eqnarray}
X_2 =&& (1+ \frac{2}{3} s^2_w) a_{z3} -4 s^2_w a_{z4}  \; ,
\nonumber \\
X_3 =&& a_{z3} + 4 c^2_w a_{z2} \; , \nonumber \\
X_4 =&& a_{z3} - \frac{4}{3} c^2_w a_{z2} \, . 
\label{X123}  
\end{eqnarray}

The anomalous amplitudes $awzt_{--}$ and $awzt_{+-}$
can be ignored in our analysis.  The reason is because the
${\cal {L}}^{(4)}$ amplitudes $wzt_{--}$
and $wzt_{+-}$ are zero, which means that,
when we consider the total helicity amplitudes squared,
they turn out to be of order $1/ {\Lambda}^2$.  This is why only
$awzt_{++}$ and $awzt_{-+}$ are presented in terms of the
parameters $X_2$, $X_3$ and $X_4$, each parameter
associated to a different partial wave.

\section{Numerical Results}
\indent\indent

\subsection{ \bf Top quark production rates from $V_L V_L$ fusions}

\indent
As discussed above,
the top quark productions from $V_L V_L$ fusion processes 
can be more sensitive to the electroweak
symmetry breaking sector than the 
longitudinal gauge boson productions from $V_L V_L$ fusions. 
In this section we shall examine the possible
increase (or decrease) of the top quark
event rates, due to the anomalous dimension 5 couplings, at the
future hadron collider LHC (a ${\rm pp}$ collider
with $\sqrt{s}=14$ TeV and  $100 \; {\rm {fb}}^{-1}$ of integrated
luminosity) and the electron linear collider LC  (an $e^{-} e^{+}$
collider with $\sqrt{s}=1.5$ TeV and $200 \; {\rm {fb}}^{-1}$ of integrated
luminosity).

To simplify our discussion, we shall assume an
approximate custodial symmetry
and make use of the helicity amplitudes given in the previous section to
compute the production rates for $t\ov t$ pairs and for single-$t$ or $\ov t$
quarks.  We shall adopt the effective-$W$ approximation method
\cite{dawson,effw},
and use the CTEQ3L  \cite{cteq3} parton distribution function with the
factorization scale chosen to be the mass of the $W$-boson.
For this study we do not intend 
to do a detailed Monte Carlo simulation for the detection of the top quark; 
therefore, we shall only impose a minimal set of cuts on the produced $t$ or 
$b$.   The rapidity of $t$ or $b$ produced from the $V_L V_L$ fusion process 
is required to be within $2$ (i.e. $|y^{t,b}|\leq 2$) and the transverse 
momentum of $t$ or $b$ is required to be at least $20$ GeV.  To validate the 
effective-$W$ approximation, we also require the invariant
mass $M_{VV}$ to be larger than $500$ GeV.

Since we are working in the high energy regime $E\gg v$, the
approximation made when we expand
the $V_L V_L \ra t\ov t\;or\;t\ov b$ scattering amplitudes in
powers of $E$ and keep the leading terms only, becomes
a very good one.  As noted in the previous section,
in all the $T_{\pm\pm}$ amplitudes, the dimension 5 operators
will only modify the constant term (S-wave) and the $\cos{\theta}$ 
(P-wave: $d^1_{0,0}$) dependence in the angular distributions of
the leading $E^3$ contributions, whereas all the $T_{\pm\mp}$
amplitudes have a $\sin{\theta}$ (P-wave: $d^1_{0,\pm 1}$)
dependence in their leading $E^2$ contributions.  Each of the
effective coefficients, $X$, $X_1$, $X_m$, $X_2$, $X_3$
and  $X_4$, parametrizes the contribution to one of the
partial waves.\footnote{In $W^{+}_L W^{-}_L\ra t\ov t$, $X_m$
contributes to both P-partial waves.}
Since contributions to different partial waves do not
interfere with each other, we can make a consistent analysis
by taking only one coefficient non-zero at a time.

\begin{figure}
\centerline{\hbox{
\psfig{figure=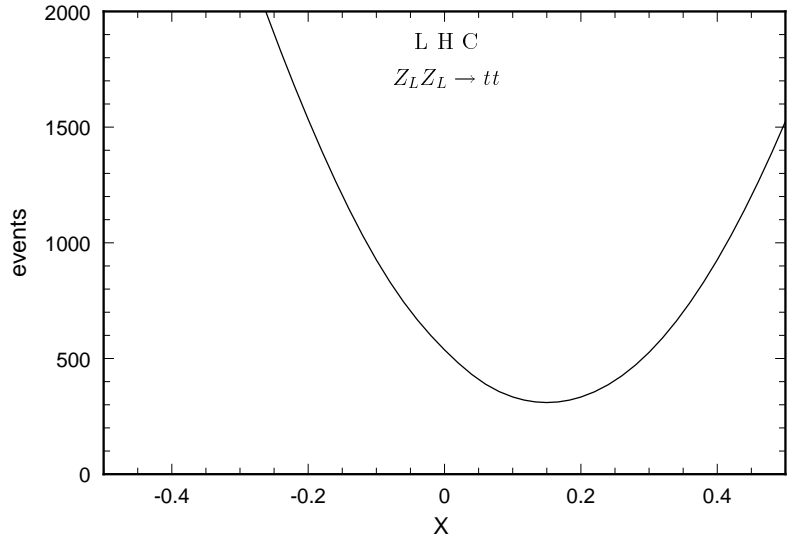,height=3in}}}
\caption{ Number of events at the LHC for  $ Z_L Z_L$
 fusion. The variable $X$ is defined in
Eq.~(\protect\ref{X}).}
\label{lhczz}
\end{figure}

\begin{figure}
\centerline{\hbox{
\psfig{figure=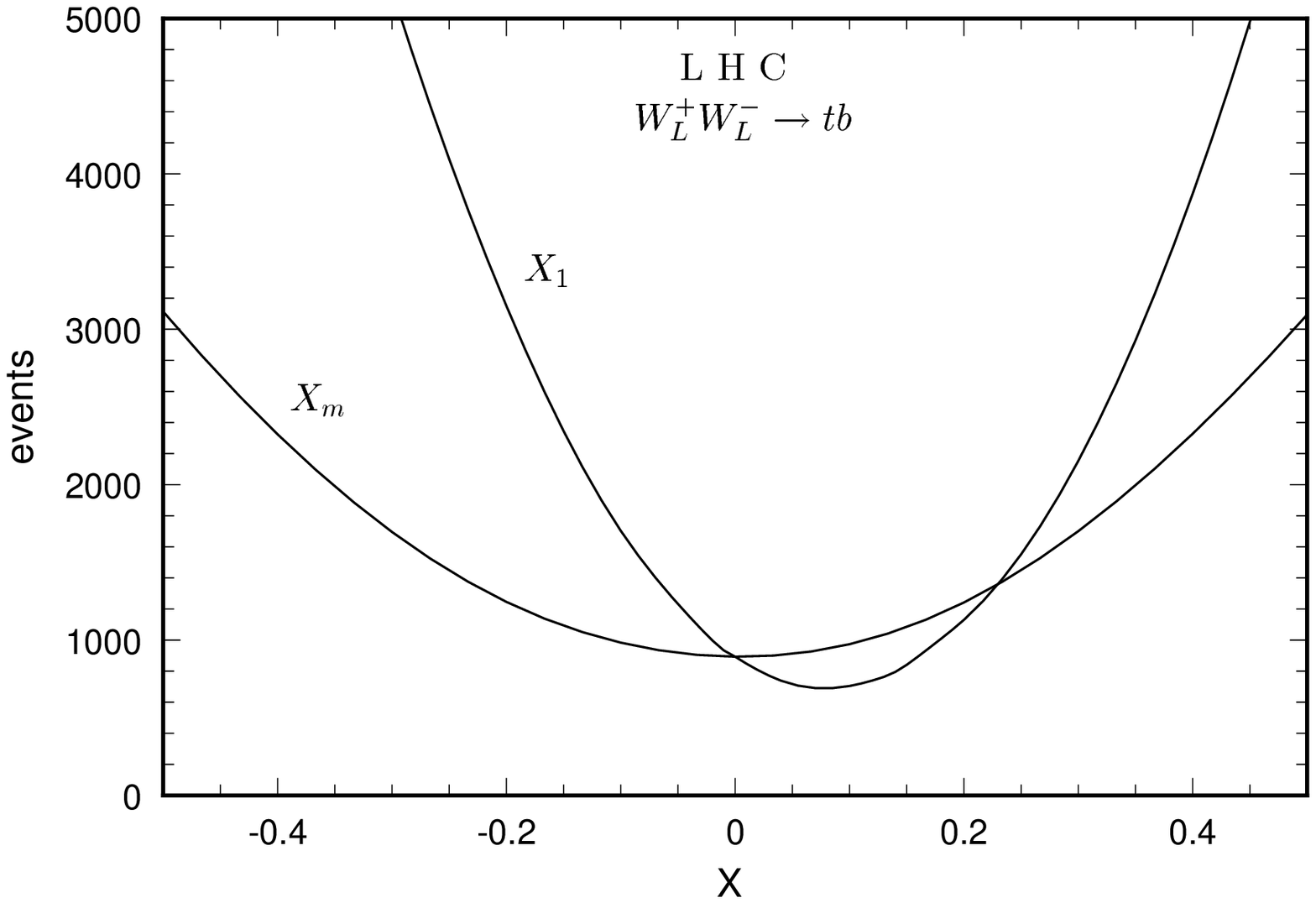,height=3in}}}
\caption{ Number of events at the LHC for  $W^{+}_L W^{-}_L$
fusion. The variable $X$ stands for the effective
coefficients $X_1$ and $X_m$ defined in
Eq.~(\protect\ref{Xprime}).}
\label{lhcww}
\end{figure}

\begin{figure}
\centerline{\hbox{
\psfig{figure=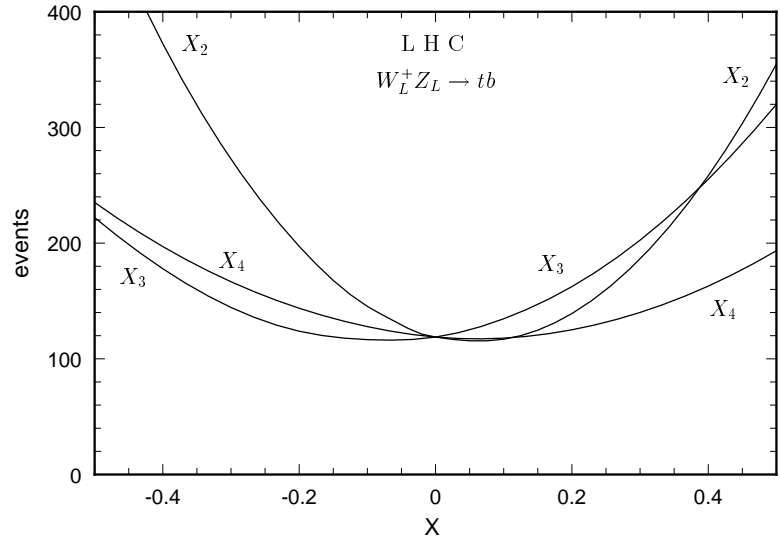,height=3in}}}
\caption{ Number of events at the LHC for $W^{+}_L Z_L$ fusion.
The variable $X$ stands for the effective
coefficients $X_2$, $X_3$ and $X_4$ defined in
Eq.~(\protect\ref{X123}).}
\label{lhcwz}
\end{figure}

\begin{figure}
\centerline{\hbox{
\psfig{figure=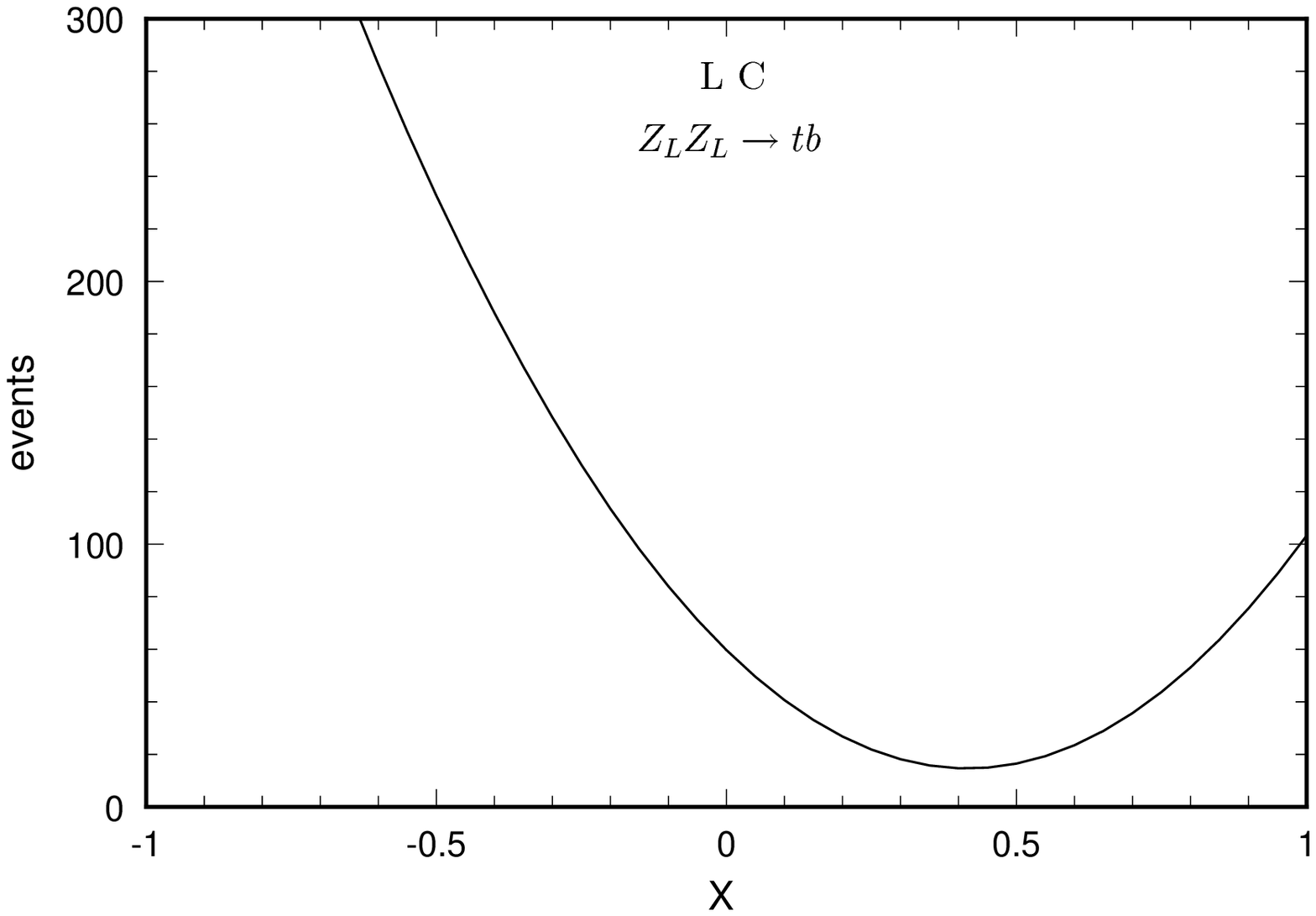,height=3in}}}
\caption{ Number of events at the LC for  $ Z_L Z_L$
 fusion. The variable $X$ is defined in
Eq.~(\protect\ref{X}).}
\label{lczz}
\end{figure}

\begin{figure}
\centerline{\hbox{
\psfig{figure=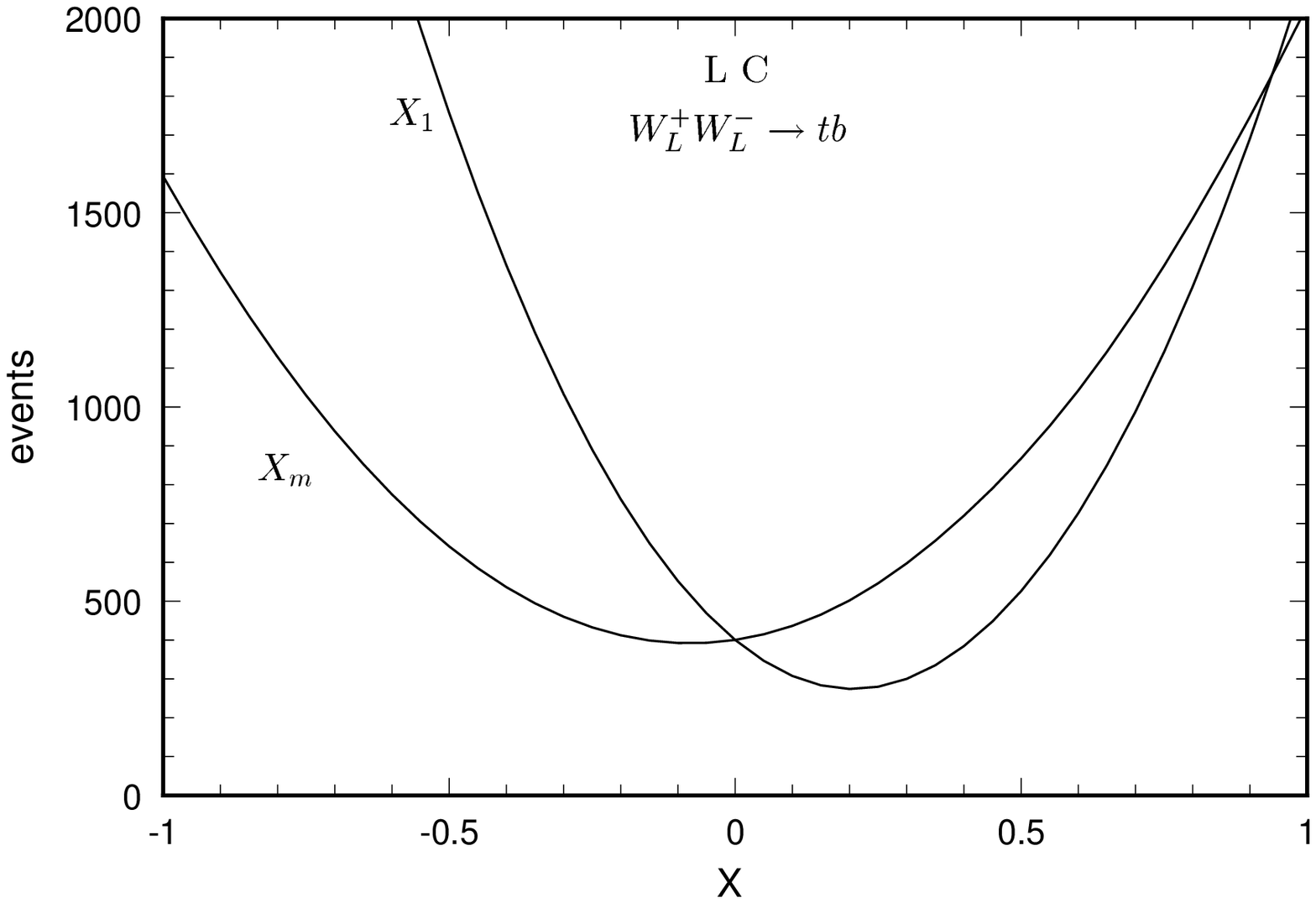,height=3in}}}
\caption{ Number of events at the LC for  $W^{+}_L W^{-}_L$
fusion. The variable $X$ stands for the effective
coefficients $X_1$ and $X_m$ defined in
Eq.~(\protect\ref{Xprime}).}
\label{lcww}
\end{figure}

\begin{figure}
\centerline{\hbox{
\psfig{figure=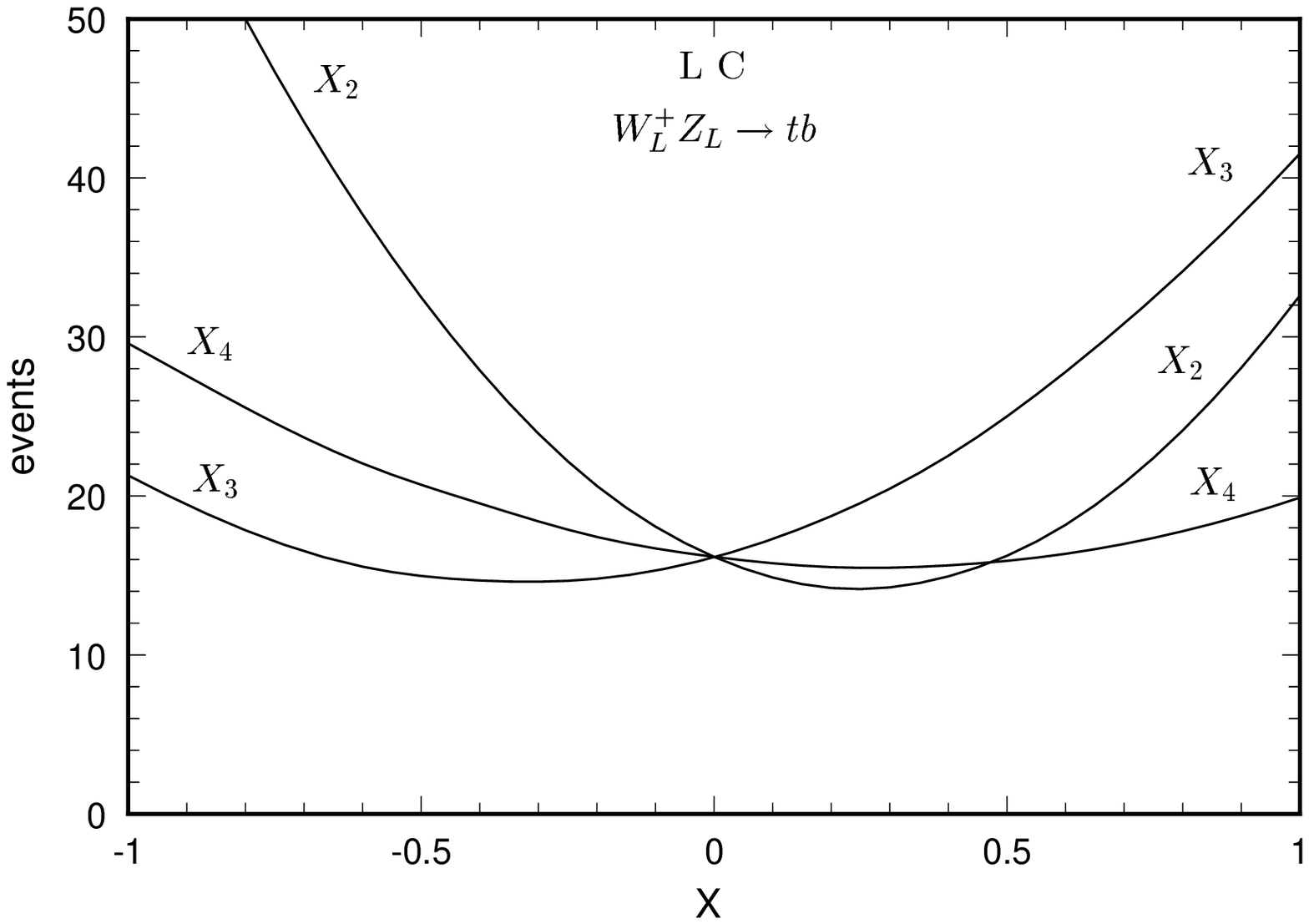,height=3in}}}
\caption{ Number of events at the LC for $W^{+}_L Z_L$ fusion.
The variable $X$ stands for the effective
coefficients $X_2$, $X_3$ and $X_4$ defined in
Eq.~(\protect\ref{X123}).}
\label{lcwz}
\end{figure}

The predicted top quark event rates as a function of these
coefficients are given in Figs. \ref{lhczz}, \ref{lhcww}
and \ref{lhcwz} for the LHC, and in Figs. \ref{lczz}, \ref{lcww}
and \ref{lcwz} for the LC.  In these plots, neither the 
branching ratio nor the detection efficiency have been included.

For $X = 0$, the LHC results show that there are in total about
1500 $t\ov t$ pair and single-$t$ or $\ov t$ events predicted by the
{\it no-Higgs} SM.   The $W^{+}_L W^{-}_L$ fusion rate is
about a factor of $2$ larger than the $Z_L Z_L$ fusion rate, and about an
order of magnitude larger than the $W^{+}_L Z_L$ fusion rate.
The $W^{-}_L Z_L$ rate, which is not shown here, is about a factor of $3$
smaller than the $W^{+}_L Z_L$ rate due to smaller parton luminosities at a 
${\rm pp}$ collider.  It will be challenging to actually detect any signal
from these channels at the LHC due to the considerable amount
of background in this hadron-hadron collision.  What we can learn from
Fig. \ref{lhcww} is that, with a production of about $900$ events and the
large slope of the $W^{+}_L W^{-}_L \ra t{\ov t}$ curve, this process
might be able to probe the anomalous coupling ($X_1$).

For the LC, because of the small coupling of $Z$-$e$-$e$,
the event rate for 
$Z_L Z_L \ra t\ov t$ is small.   For the {\it no-Higgs} SM, the top quark 
event rate at LC is about half of that at the LHC and yields a total of
about 550 events ($t\ov t$ pairs and single-$t$ or $\ov t$).
Again, we find that the $W^{+}_L W^{-}_L \ra t{\ov t}$  rate is sensitive
to the dimension 5 operators that correspond to $X_1$, 
but the $Z_L Z_L \ra t\ov t$ rate is much less sensitive.\footnote{
Needless to say, the $W^{-}_L Z_L$ rate is the same as
the $W^{+}_L Z_L$ rate at an unpolarized $e^{+}e^{-}$ LC}

The production rates shown in Figs. \ref{lczz}, \ref{lcww} and \ref{lcwz}
are for an unpolarized $e^{-}$ beam at the LC.
Because the coupling of the $W$ boson to the electron is purely
left handed, the parton luminosity of the $W$ boson will double
for a left-handed polarized $e^{-}$ beam at the LC; hence,
the $t{\ov t}$ rate from $W^{+}_L W^{-}_L $ fusion will double too.
However, this is not true for the parton luminosity of $Z$ because
in this case the $Z$-$e$-$e$ coupling is nearly purely axial-vector
($1-4s^2_w \approx 0$) and the production rate
of $Z_L Z_L \ra t\ov t$ does not strongly depend on whether
the electron beam is polarized or not.  As shown in these plots,
if the anomalous dimension 5 operators can be of
order $10^{-1}$ (as expected by the naive dimensional analysis)
then their effect can in principle be identified in the measurement of
either $Z_L Z_L$ or $W^{+}_L W^{-}_L$ fusion rates at the LC.
\footnote{
Specifically, for anomalous coefficients of order $10^{-1}$ there is
a $2\sigma$ deviation from the {\it no-Higgs} SM event rates.}
A similar conclusion holds for the $W^{\pm}_L Z_L$ fusion process,
but with less sensitivity.

From the six independent coefficients,
$a_{z(2,3,4)}$, $a_{zz1}$, $a_{ww2}$ and $a_{m}$, one stands
out: $a_{zz1}$.   The two most potentially significant parameters
$X$ and $X_1$ depend essentially on just this coefficient
[cf. Eqs. (\ref{X}) and (\ref{Xprime})].  This suggests that a
good test for the possible models of EWSB is to calculate
their predictions for the sizes of the four point operators
 $O_{g{\cal Z}{\cal Z}}$ and  $O_{g{\cal W}{\cal W}}$ because
these are more likely to produce a measurable signal at
either the LC or the LHC.    The second better test could
be the magnetic moment $a_m$ because this coefficient
gives the largest contribution to $X_m$ [cf. Eq. (\ref{Xprime})],
and Figs. \ref{lhcww} and \ref{lcww} show that this parameter
can be measured as well.

It is useful to ask for the bounds on the coefficients of the anomalous 
dimension 5 operators if the measured production rate at the LC
is found to be in agreement with the {\it no-Higgs} SM predictions
(i.e. with $X=0$).  In order to simplify this analysis for the parameter
$X_m$, we have made the approximation $aww_{+-}\simeq
{{{8E^2}\over {v^2}} m_t s_{\theta} {{X_m }\over{\Lambda}}}$;
notice that the anomalous contribution $aww_{+-}$
to the total amplitude squared is smaller by a factor of
$m_t^2/E^2$ than the contribution from $aww_{\pm \pm}$
[cf. end of section 7.2].

At the $95\%$ C.L. we summarize the bounds on the $X$'s in 
Table \ref{bounds}. 
Here, only the statistical error is included.  In practice, after including 
the branching ratios of the relevant decay modes and the detection 
efficiency of the events, 
these bounds will become somewhat weaker, but we 
do not expect an order of 
magnitude difference.   Also, these bounds shall 
be improved by carefully analyzing angular correlations when
data is available. 

\begin{table}[htbp]
\begin{center}
\vskip -0.06in
\begin{tabular}{|l||c|} \hline \hline
Process &  Bounds ( $e^{+}e^{-}$ )   \\ \hline\hline
$Z_L Z_L \rightarrow t \bar {t}$ & 
$-0.07<X<0.08$  \\ \hline
$W^{+}_L W^{-}_L \rightarrow t \bar {t}$ &   
$-0.03<X_1<0.035$  \\ \hline
$W^{+}_L W^{-}_L \rightarrow t \bar {t}$ &  
$-0.28<X_m<0.12$  \\ \hline
$W^{+(-)}_L Z_L \rightarrow t \bar {b}\; (b \bar {t})$ & 
 $-0.32 <X_2< 0.82$  \\ \hline
$W^{+(-)}_L Z_L \rightarrow t \bar {b} \;(b \bar {t})$ & 
 $-1.2 <X_3< 0.5$   \\ \hline
$W^{+(-)}_L Z_L \rightarrow t \bar {b}\; (b \bar {t})$  & 
 $-0.8 <X_4< 1.3$   \\ \hline \hline
\end{tabular}
\end{center}
\vskip 0.08in
\caption{The range of parameters for which the total number of events
at the LC deviates by less than $2\sigma$ from the {\it no-Higgs}
SM prediction.}
\label{bounds}
\end{table}

As shown in Table \ref{bounds}, these coefficients can be probed to 
about an order of  $10^{-1}$ or even $10^{-2}$.  For this Table, we have 
only considered an unpolarized $e^{-}$  beam for the LC.  To obtain the 
bounds we have set all the anomalous coefficients to be zero except the 
one of interest.  This procedure is justified by the fact that
at the leading orders of $E^3$ and $E^2$,
different coefficients contribute to different partial waves.
(The definitions of the combined coefficients $X$,
$X_1$, $X_2$, $X_3$ and $X_4$ are given in the previous section.)

If the LC is operated at the $e^{-} e^{-}$ mode with the same CM
energy of the collider, then it cannot be used to probe the effects
for $W^{+}_L W^{-}_L \rightarrow t \bar {t}$, but it can improve the
bounds on the combined coefficients $X_4$, $X_2$ and $X_3$,
because the event rate will increase by a factor of $2$ for
$W^{-}_L Z_L \ra b{\ov t}$ production.

By combining the limits on these parameters we can find the 
corresponding limits on the effective
coefficients $a_{zz1}, \, a_{z2}, \, a_{z3}, \, a_{z4}$,
and ($a_m+\frac{1}{2}{\it a_{ww2}}$).  For example, 
if we consider the limits for $X_3$ and $X_4$,
we will find the limits for $a_{z2}$, $a_{z3}$.  Then
we can compare the bounds on $a_{z3}$ and those on $X_1$
to derive the constraints on $a_{zz1}$.  Also, the
bounds on  $a_{z3}$ and on $X_2$ will give the
constraints on $a_{z4}$.  Finally, we use the bounds
on $a_{z3}$, $a_{z2}$ and $X_m$ to obtain constraints
for ($a_m+\frac{1}{2}{\it a_{ww2}}$).  Table \ref{abounds}
shows these results.

\begin{table}[htbp]
\begin{center}
\vskip -0.06in
\begin{tabular}{|l||c|} \hline \hline
Bounds on $X$ parameters &  Bounds on anomalous
coefficients   \\ \hline\hline
$-1.2 <X_3< 0.5$    &  $ -0.6 < a_{z2} < 0.32$   \\ 
 $-0.8 <X_4< 1.3$ & $-0.9<a_{z3}<1.1$ \\  \hline
$-.03<X_1<.035$ &   
$ -0.17 < a_{zz1} < 0.15$ \\ \hline
$-0.32 <X_2< 0.82$ &  $-1.9 < a_{z4} < 1.7$
 \\ \hline
$-.28<X_m<.12$ & 
$ -0.7 < a_m+\frac{1}{2}{\it a_{ww2}} < 0.4$ \\ \hline \hline
\end{tabular}
\end{center}
\vskip 0.08in
\caption{The constraints on the anomalous coefficients obtained
by the linear combination of the bounds on the $X$ parameters.}
\label{abounds}
\end{table}

Nevertheless, we can also follow the usual procedure of taking
only one anomalous coefficient as non-zero at a time.  Under this
approach the bounds become more stringent:
\begin{eqnarray}
-0.3 <&& a_{zz1}  \;\;\;\;\;\; <0.035 \; , \nonumber \\
-0.28 <&& a_m \; \;\;\;\;\;\; < 0.12 \; , \nonumber \\
-0.24 <&& a_{z3} \, \;\;\;\;\;\; < 0.28 \; , \nonumber \\
-0.4 <&& a_{z2} \, \;\;\;\;\;\; < 0.2 \; , \nonumber \\
-0.82 <&& a_{ww2} \;\;\;\; < 0.32 \; , \nonumber \\
-0.56 <&& a_{z4} \, \;\;\;\;\;\; < 0.24 \; . 
\end{eqnarray}

Again, these bounds come from the consideration of a $2 \sigma$
deviation from the {\it no-Higgs} SM event rates.  For instance, at
the LC, the {\it no-Higgs} SM predictions for the processes
$Z_L Z_L \ra t {\ov t}$ and $W^{+}_L W^{-}_L \ra t {\ov t}$ are 60
and 400, respectively [cf. Figs. \ref{lczz} and \ref{lcww}].
This means that a number of events
between 75 and 45 for the first process, and between 440 and 360
for the second one, is considered consistent with the
{\it no-Higgs} SM prediction at the $95\%$ C.L.. 
Fig. \ref{lczz} shows an interesting
situation for $Z_L Z_L \ra t {\ov t}$, if the parameter $X$ happened
to be between 0.75 and 0.90 then we would obtain a number of
events consistent with the {\it no-Higgs} SM.  However, if this
were the case, then $X_1$ would have to be at least of order 0.7
and we would observe a substantial deviation (of about 600) in the
number of events produced from $W^{+}_L W^{-}_L \ra t {\ov t}$.
This also happens the other way around, if $X_1$ is
between 0.38 and 0.45, we would obtain a production rate
consistent with the {\it no-Higgs} SM for $W^{+}_L W^{-}_L$
fusion [cf.  Fig. \ref{lcww}], but then $X$ would be at least of
order 0.3, and according to Fig. \ref{lczz}, we would observe
only 18 $t {\ov t}$ pairs from $Z_L Z_L$ fusion, too far from
the $60 \pm 15$ range of the {\it no-Higgs} SM prediction.
Hence, all the production channels have to be measured to
conclusively test the SM and probe new physics.

\begin{figure}
\centerline{\hbox{
\psfig{figure=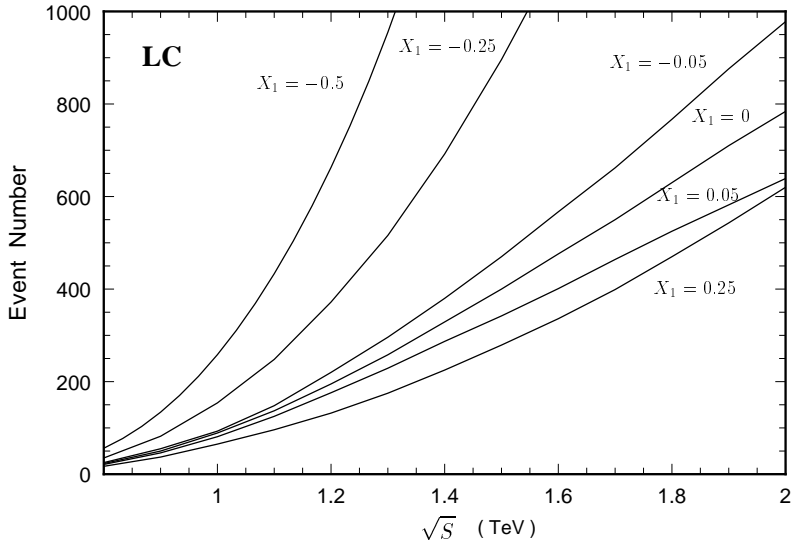,height=3in}}}
\caption{Number of $t\ov t$ events at the LC from $W^{+}_L W^{-}_L$
fusion for different values of the effective coefficient $X_1$ as a
function of the CM energy.}
\label{fwwpro}
\end{figure}

The above results are for the LC with a $1.5$ TeV CM energy.  
To study the possible new effects in the production rates of  
$W^{+}_L W^{-}_L \rightarrow t \bar {t}$ at the LC with different CM 
energies, we plot the production rates for various values of $X_1$ 
in Fig. \ref{fwwpro}. (Again, $X_1=0$ stands for the {\it no-Higgs} SM.)
Notice that, if $X_1$ were as large as $-0.5$, then a $1$ TeV  LC could
well observe the anomalous rate via $W^{+}_L W^{-}_L$ fusion.
\footnote{If 
$X_1$ is too big , partial wave unitarity can be violated at this order.}   
For $X_1\;=0.25$ the event rate at $1.5$ TeV is down by about 
a factor of $2$ from the SM event rate.\footnote{For positive values of 
$X_1$ the rate tends to diminish below the SM rate.  However, 
near $0.25$, the rate begins to rise again, toward the SM rate.}

\subsection{ \bf CP violating effects}

\indent
The complete set of anomalous dimension 5 operators listed in 
${\cal {L}}^{(5)}$ consists of operators with CP- conserving and
non-conserving parts.  In our study of the top quark production rates
we have only considered the CP-even part of these operators;
their contribution, like the one from the {\it no-Higss} SM at tree
level, is real.   However, a CP-odd operator can contribute to the
imaginary part of the helicity amplitudes, and it can only be probed
by examining CP-odd observables.

To illustrate this point, let us consider the CP-odd part of the four-point
scalar type operator $O_{g{\cal W}{\cal W}}$ and the electric dipole
moment term of $O_{{\cal A}}$ [cf. Eqs.~(\ref{ogww}) and (\ref{oa})].
After including contributions from the {\it no-Higss} SM and
from the above two CP-odd operators, 
the helicity amplitudes for the $W^{+}_L W^{-}_L \rightarrow t \bar {t}$ 
process in the $W^{+}_L W^{-}_L$ CM frame are:
\begin{eqnarray}
T_{\pm \pm} =&& \pm {{m_t E}\over {v^2}} + 
i 2 {E^3\over {v^2}} {{\left( {\tilde a}_{ww1} + 
2 a_d \, c_{\theta} \right)}\over {\Lambda}}\; , \nonumber \\
T_{+-} =&& {{2\,{m_t^2}\,{s_{\theta}}}\over 
    {\left( {{{m_b^2}}\over {2\,{E^2}}} + 
        \left( 1 - {c_{\theta}} \right) \,
\left( 1 - {{{m_t^2}}\over {2\,{E^2}}} \right)\right) {v^2}}}\; , \\ 
T_{-+} =&& 0\; , \nonumber
\end{eqnarray}
where, by $a_d$ and ${\tilde a}_{ww1}$, we refer to the imaginary part of
the coefficients of $O_{\cal A}$ and  $O_{g{\cal W}{\cal W}}$, respectively. 

One of the CP-odd observables that can measure $a_d$
and ${\tilde a}_{ww1}$ 
is the transverse polarization ($P_{\perp}$) of the top quark,
which is the degree of polarization of the top quark in the direction  
perpendicular to the plane of the
$W^{+}_L W^{-}_L \rightarrow t \bar {t}$
scattering process.  It was shown in Ref. \cite{topol} that
\begin{equation}
P_{\perp}\; =\; \frac{2 Im \left( {T_{++}^{*}T_{-+}+T_{+-}^{*} T_{--}}
\right)} {|ww_{++}|^2+|ww_{+-}|^2+|ww_{-+}|^2+|ww_{--}|^2}\;,\nonumber 
\end{equation}
which, up to the order $\frac{1}{\Lambda}$, is
\begin{equation}
P_{\perp} \;{\cong} \;
{{4 s_{\theta} E}\over {{\left( {{{m_b^2}}\over {2\,{E^2}}} + 
   \left( 1 - {c_{\theta}} \right) \,
   \left( 1 - {{{m_t^2}}\over {2\,{E^2}}} \right)  \right)}} } 
{{\left( {\tilde a}_{ww1}+2 a_d c_{\theta}\right)}\over {\Lambda}}.
\nonumber 
\end{equation}
Again, $E\,=\,\sqrt{s}$ is the CM energy of
the $W^{+}_L W^{-}_L$ system; $P_{\perp}$, by definition,
can only obtain values between $-1$ and $1$. 
For $E=1.5$ TeV, $\Lambda = 3$ TeV, and $\theta = \frac{\pi}{2}$,
or $\frac{\pi}{3}$, we obtain $P_{\perp}\;=\; 4{\tilde a}_{ww1}$,
or ${4 {\sqrt{3}}} ({\tilde a}_{ww1}+a_d$), respectively.
Since $\mid P_{\perp} \mid$ is at most $1$, this requires
$\mid {\tilde a}_{ww1}\mid  < \frac{1}{4}$ or  
$\mid {\tilde a}_{ww1}+a_d\mid  < \frac{1}{4\sqrt{3}}$.   

At a $1.5$ TeV $e^{+} e^{-}$ collider, the {\it no-Higgs} SM predicts
about 100 $t{\ov t}$ pairs, with an invariant mass between  800 GeV
and 1100 GeV, via the $W^{+}_L W^{-}_L$ fusion process .
Let us assume that $a_d=0$, and that $P_{\perp}$ can
be measured to about $\frac{1}{\sqrt{100}}=10\%$, then an
agreement between data and the {\it no-Higgs} SM prediction
($P_{\perp}=0$ at tree level) would imply
$\mid {\tilde a}_{ww1}\mid \,\leq \, 0.04$.

\section{ Conclusions}
\indent\indent

If the fermion mass generation is closely related to the EWSB mechanism,
one expects  some residual effects of this mechanism to appear 
in accordance with the mass hierarchy. 
Since the mass of top quark is heavy, it is likely that
the interactions of the top quark can deviate largely 
from the SM predictions.
In this study, we have 
applied the electroweak chiral Lagrangian to probe 
new physics beyond the SM by studying the couplings of the top quark 
to gauge bosons.  We have restricted ourselves to only consider the
interactions of the top and bottom quarks and not the flavor changing
neutral currents that involve the other light quarks.
Also, motivated by low energy data, we assume that the coupling of 
$Z$-$b$-$b$  is not strongly modified by new physics.

In section 2, we introduced the dimension 4 non-linear chiral Lagrangian
that contains the {\it no-Higgs} SM and the four unknown effective
coefficients.  Among them, two represent the non-standard couplings
associated with the left- and right-handed charged currents $\klc$ and
$\krc$, and two more for the anomalous left- and right-handed neutral
currents $\kln$ and $\krn$.    Constrains for these coefficients have
been found using precision LEP/SLC data \cite{malkawi}.
In particular, under the assumption that the \bbz vertex 
is not modified at tree level,  $\kln$ is bound within 0 and 0.15
($0.0 < \kln < 0.15$).  On the other hand,
although $\krn$ and $\klc$ are allowed to vary in the full
range of $\pm 1$, the precision LEP/SLC data do impose some
correlations among $\kln$, $\krn$, and $\klc$.  As for $\krc$, this
coupling does not contribute to the LEP/SLC
observables of interest in the limit of $m_b=0$, but it can be
constrained independently by using the CLEO measurement
of  $b\rightarrow s\gamma$:
$-0.037 \; < \; \krc \; < 0.0015$ \cite{recent,fuj}.\footnote{
We have used the updated data in Ref.~\cite{recent} and the formula in 
Ref.~\cite{fuj} to recalculate this bound.}
At the upgraded Tevatron (Run-II) and the LHC, 
$\klc$ and $\krc$ can be further tested by studying the production and the 
decay of the top quarks while $\kln$ and $\krn$ can be better measured at 
the LC.

If a strong dynamics of the electroweak symmetry breaking mechanism can
largely modify the dimension 4 anomalous couplings, it is natural to ask
whether the same dynamics can also give large dimension 5 anomalous
couplings.  In the framework 
of the electroweak chiral Lagrangian, we have
found that there are 19 independent dimension five operators 
associated with the top quark and the bottom quark system.  
Their leading contributions to the helicity amplitudes 
for $V_L V_L \ra t {\bar t}, t {\bar b}, \, {\rm or} \, b {\bar t}$
processes are given in Appendices B and C.
The high energy behavior of the above scattering processes due to the
dimension 5 operators, two powers in $E$ above the {\it no-Higgs} SM,
provides a good opportunity to test these operators on the
production of $t {\bar t}$ pairs or single-$t$ or $\ov t$ events in high
energy collisions.    Since, in the high energy 
regime, a longitudinal gauge boson is equivalent to
the corresponding would-be Goldstone boson, the
production of top quarks via $V_L V_L$ fusions shall
probe the part of the electroweak symmetry breaking sector 
which modifies the top quark interactions.
Since the dimension 4 anomalous couplings $\kappa$'s can be 
well measured at the scale of $M_Z$ or $m_t$, we expect that 
their values will be already known by the time data is available
for the study of $V_L V_L$ fusion processes in the TeV region.
Hence,  to simplify our discussion on the accuracy of the
measurement of the dimension 5 anomalous couplings at future
colliders, we have taken the dimension 4
anomalous couplings to be zero for this part of the study.  Also we have
considered a special class of new physics effects in which an underlying
custodial $SU(2)$ symmetry is assumed that gets broken in such a way as
to keep the couplings of the $Z$-$b$-$b$  unaltered. 
This approximate custodial symmetry then
relates some of the coefficients of the anomalous operators, reducing
the number of independent coefficients from 19 down to 6 only.
Then we study the contributions of these couplings to the production
rates of the top quark at the LHC and the LC. 

We find that for the leading contributions at high energies,
only the S- and P-partial wave amplitudes are modified by these
anomalous couplings. 
If the magnitudes of the coefficients of the anomalous
dimension 5 operators are allowed to be as large as $1$ 
(as suggested by the naive dimensional
analysis \cite{georgi}), then we will be able to
make an unmistakable identification of their effects to the
production rates of top quarks via the longitudinal weak boson
fusions.   However, if the measurement of the top quark 
production rate is found to agree with the SM prediction,
then one can bound these coefficients to be at most of
order $10^{-1}$.   This is about a
factor $\frac{\Lambda}{m_t} \simeq \frac{3
{\rm {TeV}}}{175 {\rm {GeV}}}\sim O(10)$ 
more stringent than in the case of the study of NLO 
bosonic operators via the $V_L V_L \ra V_L V_L$ scattering processes 
\cite{et,poc,sss}.    Hence, for those models of electroweak symmetry 
breaking for which the naive dimensional analysis gives the
correct size for the coefficients of 
dimension 5 effective operators, the top quark production
via $V_L V_L$ fusions can be a more sensitive probe to EWSB
than the longitudinal gauge boson pair production via $V_L V_L$ 
fusions which is commonly studied.  
For completeness, we also briefly discuss how to study the
CP-odd operators by measuring CP-odd observables.  
In particular, we study their effects on the transverse 
(relative to the scattering
plane of $W^{+}_L W^{-}_L \rightarrow t\bar t$) 
polarization of the top quark. 

In conclusion, the production of top quarks via $V_L V_L$
fusions at the LHC and the LC should be carefully studied when
data is available because it can be sensitive to the electroweak
symmetry breaking mechanism, even more than the commonly
studied $V_L V_L \ra V_L V_L$ processes in
some models of strong dynamics.

\medskip
\section*{ \bf  Acknowledgments }
 We thank Hong-Jian He, G. L. Kane, 
E. Malkawi and T. Tait for helpful discussions. 
F. Larios  was supported in part by the Organization of American 
States, and by the  Sistema Nacional de Investigadores.  CPY  was
supported in part by the NSF grant No. PHY-9507683.    


\newpage

\appendix

\section{Equations of Motion}

From the electroweak chiral Lagrangian ${\cal L}^{(4)}$
of Eq.~(\ref{eq2}), we can use the Euler-Lagrange equations to
obtain the equations of motion for the top quark.
They are:
\begin{eqnarray}
i{\gamma}^{\mu} ({\partial}_{\mu} +i \frac{2}{3} s^2_w {\cal A}_{\mu}) t_L- 
\frac{1}{2} (1-\frac{4}{3} s^2_w +
{\kln}) \gamma^{\mu} {\cal Z}_{\mu} t_L - 
\frac{1}{\sqrt {2}} (1+{\klc}) 
\gamma^{\mu} {\cal W}^{+}_{\mu} b_L  - m_t t_R &=& 0\, ,
\nonumber \\ 
i{\gamma}^{\mu} ({\partial}_{\mu} + 
i \frac{2}{3} s^2_w {\cal A}_{\mu}) t_R - 
\frac{1}{2} (-\frac{4}{3} s^2_w+{\krn}) \gamma^{\mu}{\cal Z}_{\mu}
t_R - \frac{1}{\sqrt {2}} {\krc} 
\gamma^{\mu} {\cal W}^{+}_{\mu} b_R  - m_t t_L &=& 0\, ,
\nonumber \\
i{\gamma}^{\mu} ({\partial}_{\mu} - i \frac{1}{3} s^2_w
{\cal A}_{\mu}) b_L - (-\frac{1}{2}+\frac{1}{3} s^2_w )
\gamma^{\mu} {\cal Z}_{\mu} b_L - \frac{1}{\sqrt {2}}
(1+{\kappa}^{CC\dagger}_{L})
\gamma^{\mu} {\cal W}^{-}_{\mu} t_L  - m_b b_R &=& 0\, ,
\nonumber \\
i{\gamma}^{\mu} ({\partial}_{\mu} - i \frac{1}{3} s^2_w
{\cal A}_{\mu}) b_R - \frac{1}{3} s^2_w  \gamma^{\mu}
{\cal Z}_{\mu} b_R - \frac{1}{\sqrt {2}}
{\kappa}^{CC\dagger}_{R} \gamma^{\mu} {\cal W}^{-}_{\mu}
t_R  - m_b b_L &=& 0\, .
\nonumber
\end{eqnarray}

\section{${\cal {L}}^{(4)}$ helicity amplitudes}

 Below, we show the leading contributions in powers of $E$ (the CM energy
of the $V_L V_L$ system) of the helicity amplitudes for the processes
$V_L V_L \ra t {\ov t}$, $t {\ov b}$ and $b {\ov t}$, in the limit 
$E \gg m_t\gg m_b$, and for the {\it no-Higgs}
SM (i.e. ${\cal {L}}^{(4)}_{SM}$).\footnote{These
amplitudes agree with those given in Ref. \cite{hinch}.} 
In general, any contribution that is not proportional to $E^3$
or $m_t E^2$ (the highest leading factors) is neglected throughout
this paper.

\subsection{ $Z_L Z_L \ra t{\ov t}$ and $W^{+}_L W^{-}_L \ra t{\ov t}$}

The helicity amplitudes for $t{\ov t}$ production are given as follows.
The first two letters, $zz$ or $ww$, refer to the 
$Z_L Z_L \ra t{\ov t}$ or $W^{+}_L W^{-}_L \ra t{\ov t}$ scattering
processes, respectively.  The first and second adjacent
symbols ($+$ or $-$), refer to
the helicities of the final top and anti-top quarks, respectively.
Throughout this paper, the scattering angle $\theta$ is defined
as the one subtended between the momentum of the incoming
gauge boson that appears on the top-left part
of the Feynman diagram [cf. Figs. \ref{fzz}, \ref{fww} and \ref{fwz}]
and the momentum of the outgoing fermion appearing on the top-right
part of the same diagram; all in the CM frame of the
$V_L V_L$ pair.  We denote its sine and cosine functions as
$s_{\theta}$ and $c_{\theta}$, respectively.
\begin{eqnarray}
zz_{++} =&&-zz_{--}\,\,=\,\, {{{m_t}\,E}\over {{v^2}}}
\; , \nonumber \\
zz_{+-} =&&zz_{-+}\,\,=\,\, {{2\,\,{{{m^2_t}}}\,
{c_{\theta}} {s_{\theta}}}\over 
    {\left( {{4{{{c^2_{\theta}}}}\, m^2_t}\over {{E^2}}} + 
   {{{s^2_{\theta}}}} \right) \,{v^2}}} \; , \nonumber\\
ww_{++}=&&-ww_{--}\;\;\;=\;\;\;zz_{++} \; ,\nonumber \\
ww_{+-} =&& {{2\,{{{m^2_t}}}\,{s_{\theta}}}\over 
    {\left( {2{{{{m_b}}^2}}\over {{E^2}}} + 
        \left( 1 - {c_{\theta}} \right) \,
   \left( 1 - {2{{{{m^2_t}}}}\over {{E^2}}} \right)  \right) \,
   {v^2}}} \; , \nonumber\\
ww_{-+}=&& {{2\,{{{m^2_b}}}\,{s_{\theta}}}\over 
    {\left( {2{{{{m_b}}^2}}\over {{E^2}}} + 
        \left( 1 - {c_{\theta}} \right) \,
   \left( 1 - {2{{{{m^2_t}}}}\over {{E^2}}} \right)  \right) \,
   {v^2}}}  \; . \hspace{6cm} (B.1) \nonumber
\end{eqnarray}
For $ww_{+-}$ we have kept the term proportional to the $b$-mass
in the denominator of the fermion-propagator to avoid infinities
at ${\theta}=0$ in the numerical computations.

For completeness, we include the leading contributions that may 
come from the $\kappa$ coefficients in ${\cal {L}}^{(4)}$
[cf. Eq.~(\ref{eq2})]:
\begin{eqnarray}
zz^{\kappa}_{++} =&&-zz^{\kappa}_{--}\,\,=\,\, 
{{\,{m_t}\, E }\over {{v^2}}}
\left[\, (\kln-\krn+1)^2-1\, \right] \; , \nonumber \\
zz^{\kappa}_{+-} =&&zz^{\kappa}_{-+}\,\,=\,\, 
{{2\, {m^2_t}\, {c_{\theta}} {s_{\theta}}}\over 
{\left( {4{{c^2_{\theta}}\,{m^2_t}}\over {{E^2}}} + 
 {{{s^2_{\theta}}}} \right) \,{v^2}}}
\left[\, (\kln-\krn+1)^2-1\, \right] \; , \nonumber \\
ww^{\kappa}_{++}=&& {{{m_t}\, E }\over {{v^2}}}
\left[{(1+c_{\theta}) (\,2{\klc}+({\klc})^2+({\krc})^2\,)-
c_{\theta}({\kln}+{\krn})}\right] \, , \nonumber\\
ww^{\kappa}_{--} =&& -ww^{\kappa}_{++} \nonumber\\
ww^{\kappa}_{+-} =&& \frac{E^2 s_{\theta}}{v^2}
\left[\,{\krn}-({\krc})^2\,\right] \; , \nonumber \\
ww^{\kappa}_{-+}=&&\frac{E^2 s_{\theta}}{v^2}
\left[{{\kln}-{\klc}(2+{\klc})}\right] \; . \hspace{6.5cm} (B.2)
\nonumber
\end{eqnarray}

\subsection{$W^{+}_L Z_L \ra t{\ov b}$ and $W^{-}_L Z_L \ra b{\ov t}$}

The following helicity amplitudes for single top or anti-top production were
not given in Ref. \cite{hinch}.  We have taken the limit $E\gg m_t\gg m_b$.
The first three letters, $wzt$ or $wzb$, refer to the 
$W^{+}_L Z_L \ra t{\ov b}$ or $W^{-}_L Z_L \ra b{\ov t}$ scattering 
process, respectively.  
\begin{eqnarray}
wzt_{++} =&& -{{{\sqrt{2}}\,{m^3_t} \,
\left( 1 - {c_{\theta}} \right) }\over 
{E\,{{\left( 1 - {2{{{{\it m^2_t}}}}\over {{E^2}}} \right) }}\,
\left( 1 + {c_{\theta}} + {2{{{{\it m^2_t}}}}\over 
           {{E^2}}} \right) \,{v^2}}} \; , \nonumber\\
wzt_{--}=&&0 \; , \nonumber \\ 
wzt_{+-}=&&0 \; , \nonumber \\
wzt_{-+} =&& -{{{\sqrt{2}}\,{{{m^2_t}}}\,{s_{\theta}}}\over 
  {{{\left( 1 - {2{{{m^2_t}}}\over {{E^2}}} \right) }}\,
   \left( 1 + {c_{\theta}} + {2{{{{m^2_t}}}}\over 
 {{E^2} }} \right) v^2 }} \; , \nonumber \\
wzb_{++}=&&-wzt_{--}(c_{\theta} \rightarrow -c_{\theta})\, \,=\,\,0
\; , \nonumber\\
wzb_{--}=&&-wzt_{++}(c_{\theta} \rightarrow -c_{\theta}) \,\,=\,\,
 {{{\sqrt{2}}\,{m^3_t} \, \left( 1 + {c_{\theta}} \right) }\over
{E\,{{\left( 1 - {2{{{{m^2_t}}}}\over {{E^2}}} \right)}}\,
\left(1 - {c_{\theta}} + {2{{{m^2_t}}}\over {{E^2}}} \right) \,{v^2}}}\; ,
\nonumber \\
wzb_{-+}=&&-wzt_{-+}(c_{\theta} \rightarrow -c_{\theta}) \,\,=\,\,
{{{\sqrt{2}}\,{{{m^2_t}}}\,{s_{\theta}}}\over 
  {{{\left( 1 - {2{{{{m^2_t}}}}\over {{E^2}}} \right) }}\,
  \left( 1 - {c_{\theta}} + {2{{{{m^2_t}}}}\over 
 {{E^2}}} \right) v^2}}\; , \nonumber\\
wzb_{+-}=&&-wzt_{+-} (c_{\theta} \rightarrow -c_{\theta})\,\,=\,\,0
\; .  \hspace{6cm} (B.3) \nonumber
\end{eqnarray}
Including the contributions from the $\kappa$ coefficients in
${\cal {L}}^{(4)}$, we obtain:
\begin{eqnarray}
wzt^{\kappa}_{++} =&&  \frac{E\,m_t}{v^2 \sqrt{2}}
(1+{\klc})\, \left[\, (1-c_{\theta}) {\kln}-2{\krn} \,\right]
\; , \nonumber \\
wzt^{\kappa}_{--} =&&  \frac{E\,m_t}{v^2 \sqrt{2}}
{\krc}\; \left[\,2{\kln}+ (1-c_{\theta})(2-{\krn}) \, \right]
\; , \nonumber \\
wzt^{\kappa}_{+-} =&&  \frac{E^2 s_{\theta}}{v^2 \sqrt{2}}
{\krc}\; (\,{\krn}-2 \,)
\; , \nonumber \\
wzt^{\kappa}_{-+} =&&  \frac{E^2 s_{\theta}}{v^2 \sqrt{2}}
{\kln}\; (\, 1+{\klc} \,) \; . \hspace{6.5cm} (B.4) \nonumber
\end{eqnarray}

\section{${\cal {L}}^{(5)}$ helicity amplitudes}

Below, we show the anomalous coupling contributions to the
helicity amplitudes for the $V_L V_L \ra t{\ov t}$, $t{\ov b}$
or $b{\ov t}$ scattering processes. The first letter, $a$, stands
for {\it anomalous}.  All the 19 anomalous operators
listed in section 4 have been considered.

\subsection{ $Z_L Z_L \ra t{\ov t}$}

There are four operators relevant to this process.
The four-point operator $O_{g{\cal Z}{\cal Z}}$, with coefficient
$a_{zz1}$, contributes only through the diagram of Fig. \ref{fzz}(c).
The other three, $O_{\sigma D {\cal Z}}$, $O_{{\cal Z}D f }$
and $O_{g D {\cal Z}}$, with coefficients $a_{z2}$, $a_{z3}$
and $a_{z4}$, respectively, contribute through diagrams \ref{fzz}(a)
and \ref{fzz}(b).  However, since external on-shell $Z$ bosons
satisfy the condition $p_\mu \epsilon^\mu = 0$,
the contribution from the derivative-on-boson operators
$O_{\sigma D {\cal Z}}$ and  $O_{g D {\cal Z}}$ vanishes.
The only non-zero contributions come from
$O_{g{\cal Z}{\cal Z}}$ and $O_{{\cal Z}D f }$.
The anomalous contributions to the helicity amplitudes are:
\begin{eqnarray}
azz_{++} =&& {{-{E^3}\,\left( 2\,{a_{zz1}} + 
 \,\left(1 - \frac{8}{3} {s^2_w}  \right) {a_{z3}} \, \right) }
     \over {{v^2 \Lambda}}} \; , \nonumber\\
azz_{--}=&&-azz_{++} \; ,  \nonumber \\
azz_{+-}=&&az_{-+} \, = \, 0 \; ,\nonumber \hspace{8.5cm} (C.1)
\end{eqnarray}
The amplitudes with opposite sign helicities
$azz_{+-}$ and $azz_{-+}$ appear as zero.  This is so because the
contribution from the four-point operator $O_{g{\cal Z}{\cal Z}}$
is proportional to the spinor product
${\ov u}[\lambda =\pm1] v [\lambda =\mp1]$, which is zero
in the CM frame of the $t {\ov t}$ pair.  Furthermore, for the operator with
derivative-on-fermion, $O_{{\cal Z}D f }$, the leading energy power
for $azz_{\pm \mp}$ is $E^0$ and we do not include it in the above
results.

\subsection{$W^{+}_L W^{-}_L \ra t{\ov t}$}

The relevant operators are:
the four-point operators $O_{g{\cal W}{\cal W}}$
and $O_{\sigma {\cal W}{\cal W}}$ with coefficients $a_{ww1}$
and $a_{ww2}$, respectively; derivative-on-boson operators
$O_{\sigma D {\cal Z}}$, $O_{g D {\cal Z}}$,
$O_{\sigma D {\cal W} L(R)}$, $O_{g D {\cal W} L(R)}$ and
$O_{ {\cal A}}$, with coefficients $a_{z2}$, $a_{z4}$,
$a_{w2 L(R)}$, $a_{w4 L(R)}$ and $a_{m}$, respectively;
derivative-on-fermion operators $O_{{\cal Z}D f }$,
$O_{{\cal W}D t R (L)}$ and $O_{{\cal W}D b L (R)}$, with coefficients 
$a_{z3}$, $a_{w3R(L)}$ and $a_{bw3L(R)}$, respectively.

However, some operators give null contributions.
For instance, $a_{w2 L(R)}$ and $a_{w4 L(R)}$ enter in the
{\it t-channel} diagram of Fig. \ref{fww}(a), but the condition
${\epsilon}_{\mu} p^{\mu} = 0$ for the on-shell $W^{+}$
and $W^{-}$ bosons makes their contribution to vanish.
Similarly, the contribution from $O_{g{\cal W}{\cal W}}$ is
proportional to the spinor product ${\ov u}[\lambda =\pm1]
v [\lambda =\mp1]$, which is zero in the $t {\ov t}$ CM frame; also,
the contribution from  $O_{g D {\cal Z}}$, which enters in the
{\it s-channel} diagram \ref{fww}(c), vanishes when the Lorentz
contraction in the product of the tri-boson coupling, the bosonic
propagator and the anomalous coupling is done.
There is no effect from operators that depend on $b_R$, such as
$O_{gD{\cal W}L}$,  $O_{{\cal W}D t L}$ and  $O_{{\cal W}D b L}$,
because the bottom quark is purely left handed in diagram \ref{fww}(a)
in the limit $m_b\ra 0$. Also, the contributions from the operators
$O_{{\cal W}D t R (L)}$ (with coefficient $a_{w3R(L)}$) and  
$O_{{\cal W}D b L (R)}$ (with coefficient $a_{bw3L(R)}$)
are identical.  Hence, the helicity amplitudes are:
\begin{eqnarray}
aww_{++} =&& {-\, {{2 E^3}\over {v^2 \Lambda}} \left( {\it a_{ww1}}
+ {a_{ww2}}\,{c_{\theta}} \right) }\; -  \nonumber \\ 
  && {{E^3}\over {v^2 \Lambda}}\left(
{{{a_{w3R}+a_{bw3R}}}\over {{\sqrt{2}}}} +   
   { c_{\theta}}\,\left( -{{{a_{w3R}+a_{bw3R}}}\over {{\sqrt{2}}}} - 2\,
{a_{z2}} +{a_{z3}} +  4 a_m \right) \right) \; , \nonumber \\
aww_{--}=&&-aww_{++} \; ,  \nonumber \\
aww_{+-} =&& {{{2\,{E^2}\,{m_t}\,{s_{\theta}}\over {v^2 \Lambda}}\,
 \left( 2\,{a_{ww2}} - {{{a_{w3R}+a_{bw3R}}}\over {{\sqrt{2}}}} -
 2\,{a_{z2}} + 4 a_m \right) }} \; , \nonumber\\
aww_{-+} =&& {{{2\,{E^2}\,{m_t}\,{s_{\theta}}\over {v^2 \Lambda}}\,
 \left( 2\,{a_{ww2}} - 2\,{a_{z2}}  + 4 a_m \right) }} \; .
\hspace{5cm} (C.2) \nonumber
\end{eqnarray}

\subsection{ $W^{+}_L Z_L \ra t{\ov b}$}

There are two kinds of operators that contribute to this process.
The first ones (operators with top and bottom quarks) distinguish
chirality; the second ones (operators with top quarks only) do not.
The ones that distinguish chirality are: the four-point operators
$O_{g{\cal W}{\cal Z} L (R)}$ and $O_{\sigma {\cal W}{\cal Z} L (R)}$,
with coefficients $a_{wz1 L (R)}$ and $a_{wz2 L (R)}$, respectively;
derivative-on-boson operators $O_{\sigma D {\cal W} L(R)}$ and
$O_{g D {\cal W} L(R)}$, with coefficients $a_{w2 L(R)}$ and
$a_{w4 L(R)}$, respectively; derivative-on-fermion operators
$O_{{\cal W}D t R (L)}$ and $O_{{\cal W}D b L (R)}$, with
coefficients $a_{w3R(L)}$ and $a_{bw3L(R)}$, respectively.
The second ones, that do not distinguish chirality, are:
derivative-on-boson operators $O_{\sigma D {\cal Z}}$ and
$O_{g D {\cal Z}}$, with coefficients $a_{z2}$ and $a_{z4}$,
respectively; derivative-on-fermion operator $O_{{\cal Z}D f }$,
with coefficient $a_{z3}$.

A particular feature, common to all the operators that distinguish
chirality, takes place:
If the helicity of the particle is {\it opposite} to the chirality in
the coupling, then the contribution will be proportional to the mass
of that particle.  For instance, the leading term for the contribution
of $O_{{\cal W}D t R }$ to $awzt_{++}$ is proportional to $E^3$,
but the leading term for $awzt_{--}$ is proportional to $m_t m_b E^1$.
(The left handed helicity of the anti-bottom is {\it opposite} to its
left handed chiral component.) 

The three relevant operators that do not distinguish chirality
participate only through the {\it u-channel} diagram of
Fig. \ref{fwz}(b), and only $O_{{\cal Z}D f }$ gives non-zero
contribution.  The other two, with derivative on boson,
have their contribution vanished from the condition
${\epsilon}_{\mu} p^{\mu} = 0$ of the on-shell $Z$ boson.
On the other hand, the contribution of $O_{{\cal Z}D f }$
to those amplitudes with a left handed helicity anti-bottom is zero
in the limit $m_b \ra 0$ because the bottom becomes purely
left handed in this diagram.  Hence,
\begin{eqnarray}
awzt_{++} =&&{{ {E^3}}\over {2v^2 \Lambda}}\, \left(
{a_{w3R}+a_{bw3R}} \left( 1 +  \frac{2}{3} {s^2_w} +
c_{\theta} \right) \right. \, -  \nonumber \\
 && \left. 4\,{\it a_{wz1R}} -  4\,{\it a_{wz2R}}\,{c_{\theta}} - 
    {\sqrt{2}}\,{a_{z3}}\,\left(1 + {c_{\theta}} \right)  - 
  4 {c^2_w} {a_{w2R}}\,{c_{\theta}}\, 
+ 4 {s^2_w} a_{w4R}  \, \right) \; , \nonumber \\
awzt_{+-} =&& {{{E^2} {m_t} \, {s_{\theta}}}\over {2v^2 \Lambda}}\,
 \left( 4\,{a_{wz2L}} + {a_{w3L}+a_{bw3L}} + 4 {c^2_w}\,{a_{w2L}}\, 
\right)  \; , \nonumber \\
awzt_{-+} =&& {{ {E^2} {m_t} \, {s_{\theta}}}\over {2v^2 \Lambda}}\,
\left( 4\,{a_{wz2R}} - {a_{w3R}+a_{bw3R}} - {\sqrt{2}}\,{a_{z3}} +
 4 {c^2_w} \, {a_{w2R}}\, \right) \; , \nonumber\\
awzt_{--} =&& {{{E^3}}\over {2v^2 \Lambda}}\,\left( 4\,{a_{wz1L}} + 
4\,{a_{wz2L}}\,{c_{\theta}} + 4 {c^2_w} \,
{a_{w2L}}\,{c_{\theta}}\, + \right.  \nonumber \\
&& \left. {a_{w3L}+a_{bw3L}}\,(1  - \frac{2}{3} {s_w^2} - c_{\theta} )
 -4s^2_w a_{w4L}\, \; \right) \; ,\hspace{3.5cm} (C.3) \nonumber
\end{eqnarray}

\subsection{$W^{-}_L Z_L \ra b{\ov t}$}

This process is similar to $W^{+}_L Z_L \ra t{\ov b}$, as discussed above.
The same kind of operators contribute here, and the same reasons
of why some contributions are negligible or zero apply.
\begin{eqnarray}
awzb_{++} =&& -awzt_{--} \left( c_{\theta} \rightarrow -
c_{\theta} \right)  \nonumber\\
=&& {{{E^3}}\over {2v^2 \Lambda}}\,\left( -4\,{a_{wz1L}} + 4\,
{a_{wz2L}}\,{c_{\theta}} + 4 c^2_w \,{a_{w2L}}\,
{c_{\theta}}\, - \right.   \nonumber \\
&&\left. {a_{w3L}+a_{bw3L}}\, (1 - \frac{2}{3} {s_w^2} +
{c_{\theta}}) + 4  s^2_w  a_{w4L}  \; \right) \; , \nonumber \\
awzb_{--} =&& -awzt_{++} \left( c_{\theta} \rightarrow -c_{\theta} 
\right) \nonumber\\
=&& {{{E^3}}\over {2v^2 \Lambda}}\,\left( 4\,{a_{wz1R}} + 
 {\sqrt{2}}\,{a_{z3}} \left( 1 - {c_{\theta}} \right)\, - 
\right. \nonumber \\ 
&& \left. 4\,{a_{wz2R}}\,{c_{\theta}} -  
4 c^2_w \,{a_{w2R}}\,{c_{\theta}}\, - 
 {a_{w3R}+a_{bw3R}}\,(1 + {\frac{2}{3}{{s_w}^2}} - c_{\theta} )
- 4 s^2_w  a_{w4R}  \;\right) \; ,  \nonumber\\
awzb_{+-} =&& -awzt_{+-} \; , \nonumber \\
awzb_{-+} =&& -awzt_{-+} \; . \hspace{9.5cm} (C.4) \nonumber
\end{eqnarray}


\end{document}